\documentclass[acmlarge]{acmart}

\usepackage{algorithmic}
\usepackage{graphicx}
\usepackage{textcomp}
\usepackage{threeparttable}
\usepackage{comment}
\usepackage{units}
\usepackage[binary-units=true]{siunitx}
\newcommand{\rev}[1]{#1}
\newcommand{\minor}[1]{#1}
\AtBeginDocument{%
  \providecommand\BibTeX{{%
    \normalfont B\kern-0.5em{\scshape i\kern-0.25em b}\kern-0.8em\TeX}}}

\setcopyright{acmcopyright}
\acmJournal{HEALTH}
\acmYear{2021} \acmVolume{1} \acmNumber{1} \acmArticle{1} \acmMonth{9} \acmPrice{15.00}\acmDOI{10.1145/3487910}

\begin{document}

\title{Embedding Temporal Convolutional Networks for Energy-Efficient PPG-Based Heart Rate Monitoring}

\author{Alessio~Burrello}
\affiliation{%
  \institution{Department of Electrical, Electronic and Information Engineering, University of Bologna}
  \city{Bologna}
  \state{Emilia-Romagna}
  \country{Italy}
  \postcode{40136}
}
\email{name.surname@unibo.it}

\author{Daniele~Jahier~Pagliari}
\affiliation{%
  \institution{Department of Control and Computer Engineering at Politecnico di Torino}
  \city{Torino}
  \country{Italy}
  \postcode{10129}}
\email{name.surname@polito.it}

\author{Pierangelo~Maria~Rapa}
\affiliation{%
  \institution{Department of Electrical, Electronic and Information Engineering, University of Bologna}
  \city{Bologna}
  \state{Emilia-Romagna}
  \country{Italy}
  \postcode{40136}
}
\email{name.surname@studio.unibo.it}

\author{Matilde~Semilia}
\affiliation{%
  \institution{Department of Electrical, Electronic and Information Engineering, University of Bologna}
  \city{Bologna}
  \state{Emilia-Romagna}
  \country{Italy}
  \postcode{40136}
}
\email{name.surname@studio.unibo.it}

\author{Matteo~Risso}
\affiliation{%
  \institution{Department of Control and Computer Engineering at Politecnico di Torino}
  \city{Torino}
  \country{Italy}
  \postcode{10129}}
\email{name.surname@studenti.polito.it}

\author{Tommaso~Polonelli}
\affiliation{%
  \institution{PBL Center}
  \city{Zurich}
  \country{Switzerland}
  \postcode{8092}}
\email{tommaso.polonelli@pbl.ee.ethz.ch}

\author{Massimo~Poncino}
\affiliation{%
  \institution{Department of Control and Computer Engineering at Politecnico di Torino}
  \city{Torino}
  \country{Italy}
  \postcode{10129}}
\email{name.surname@polito.it}

\author{Luca~Benini}
\affiliation{%
  \institution{Department of Information Technology and Electrical Engineering at the ETH Zurich}
  \city{Zurich}
  \country{Switzerland}
  \postcode{8092}
}
\email{lbenini@iis.ee.ethz.ch}

\author{Simone~Benatti}
\affiliation{%
  \institution{Department of Electrical, Electronic and Information Engineering, University of Bologna}
  \city{Bologna}
  \state{Emilia-Romagna}
  \country{Italy}
  \postcode{40136}
  \institution{ \newline Department of Sciences and Methods for Engineering, University of Modena e Reggio Emilia}
  \city{Modena}
  \state{Emilia-Romagna}
  \country{Italy}
  \postcode{42123}
}
\email{name.surname@unibo.it}
\renewcommand{\shortauthors}{Alessio Burrello, et al.}


\begin{abstract}
\rev{Photoplethysmography (PPG) sensors allow for non-invasive and comfortable heart-rate (HR) monitoring, suitable for compact wrist-worn devices. Unfortunately, Motion Artifacts (MAs) severely impact the monitoring accuracy, causing high variability in the skin-to-sensor interface. Several data fusion techniques have been introduced to cope with this problem, based on combining PPG signals with inertial sensor data. Until know, both commercial and reasearch solutions are computationally efficient but not very robust, or strongly dependent on hand-tuned parameters, which leads to poor generalization performance.
In this work, we tackle these limitations by proposing a computationally lightweight yet robust deep learning-based approach for PPG-based HR estimation. Specifically, we derive a diverse set of Temporal Convolutional Networks (TCN) for HR estimation, leveraging Neural Architecture Search (NAS). Moreover, we also introduce ActPPG, an adaptive algorithm that selects among multiple HR estimators depending on the amount of MAs, to improve energy efficiency. We validate our approaches on two benchmark datasets, achieving as low as 3.84 Beats per Minute (BPM) of Mean Absolute Error (MAE) on PPGDalia, which outperforms the previous state-of-the-art. Moreover, we deploy our models on a low-power commercial microcontroller (STM32L4), obtaining a rich set of Pareto optimal solutions in the complexity vs. accuracy space.}
\end{abstract}



\maketitle

\section{Introduction}\label{sec:introduction}
Wrist-worn devices are gaining remarkable traction in the wearable ecosystem~\cite{nelson2020guidelines}, benefiting from the availability of accurate, compact sensors and energy-efficient microcontrollers~\cite{al2018review}. While first-generation platforms relied mostly on accelerometers to perform activity recognition tasks, the novel paradigms of personalized healthcare and medical Internet of Things (IoT)~\cite{yeole2016use} push in the direction of having wearables with richer sensor sets to monitor vital parameters such as electrodermal activity (EDA), and heart rate (HR).

In particular, monitoring HR and HR variability is paramount for clinical purposes and precise activity monitoring. The first wrist-worn HR tracking devices were connected to a chest band with a simple 1-3 leads Electrocardiogram (ECG) sensor. Albeit accurate, this solution is uncomfortable for users and even impossible to wear in certain conditions.
More recently, the optimization and miniaturization of photoplethysmogram (PPG) sensors allowed integrating HR and blood oxygenation (SpO2) in wrist-worn devices, resulting in a more comfortable and user-friendly solution at a lower cost compared to conventional ECG strips. PPG sensors and HR estimation are now integrated on commercial wrist-worn devices such as the Apple Watch~\cite{applewatch} or the Fitbit Charge 4~\cite{fitbit}.
%

%
A major challenge in PPG-based HR estimation \cite{castaneda2018review} is represented by motion artifacts (MA), which cause variability of sensor pressure on the skin or ambient light leaking into the gap between the photodiode and the skin. Besides, blood flow can vary considerably as the type of physical activity varies, contributing to a worsening of the light absorption measurement, which is necessary to estimate HR. Several studies compared ECG and PPG based approaches, evaluating the HR on 50 \cite{ge2016evaluating} and 30 \cite{jan2019evaluation} healthy subjects in three scenarios (i.e., sitting, walking, and jogging). These studies have shown that ECG chest straps outperform PPG-based platforms, affected by a mean error of up to 10 beats per minute (BPM). As a result, the ECG-based solutions are still considered the reference benchmark for wearable HR tracking~\cite{neurosky}.

To overcome these accuracy limitations, recent research has focused on obtaining precise HR measurements by integrating PPG and accelerometer data to mitigate MAs' effect.
These studies have explored novel algorithmic approaches and collected vast datasets to validate them~\cite{troika2014, reiss2019deep}.
On the algorithmic side, most solutions are based on classical approaches such as Independent Component Analysis (ICA), Kalman Filters, and Wavelet decomposition. These methods reduce the noise caused by MAs by creating models for the PPG signal and for the noise. TROIKA \cite{troika2014} and its evolutions, JOSS \cite{joss2015} and more recently CurToSS \cite{zhou2020heart}, are seminal works in this category. They estimate the noise via adaptive filtering and then apply a spectral peak tracking to detect the heartbeat frequency achieving a Mean Absolute Error (MAE) lower than 2 beats per minute. A major shortcoming of the aforementioned approaches is related to the extensive hand-tuning of the model parameters, leading to lack of generalization.


\rev{Recently, deep learning approaches are receiving increasing attention also in this field, reaching accuracies comparable to those of classical methods with less hand-tuning of parameters. However, the deployment of these computationally-expensive models on resource-constrained platforms (such as those embedded in wrist-worn wearables) is still an open challenge since coupling a high accuracy with a small model is not trivial.
In addition to that, data-driven algorithms necessitate a high amount of training data to reach satisfactory results, which are not trivial to collect.}
\rev{In 2019, a novel PPG-based HR tracking dataset called PPGDalia was presented~\cite{reiss2019deep}, which covers many daily activities for 15 healthy subjects, paving the way to a more in-depth exploration of deep-learning solutions for this challenge.}

\rev{In this paper, which extends the preliminary work of \cite{Risso2021}, we introduce a rich set of accurate yet computationally efficient Temporal Convolutional Networks (TCNs) for PPG-based HR estimation.
Moreover, we integrate these TCNs in a novel framework, which adaptively combines multiple HR tracking models at runtime depending on an estimate of the MAs to reduce the overall energy consumption.
In detail, the main contributions of this work are the following:}
\begin{itemize}
    \item \rev{\emph{TimePPG}, a collection of Pareto-optimal TCNs for HR estimation based on raw PPG and acceleration data. All the TCNs are automatically derived from a single seed architecture~\cite{zanghieri2019robust} using a Neural Architecture Search (NAS) algorithm~\cite{gordon2018morphnet} to explore the accuracy vs. model size and the accuracy vs. complexity trade-offs.}
    \item \rev{\emph{ActPPG}, a new framework to adaptively combine TimePPG solutions, as well as other algorithms, depending on the user's movement conditions, using a bigger but more accurate model when higher MAs are expected. To this end, ActPPG employs a low-cost human activity recognition (HAR) model and uses the predicted activity as a discriminator to choose between different HR tracking models.}
    \item \rev{A detailed comparison of our approach with state-of-the-art methods, including both model-driven approaches and deep-learning ones, on two popular benchmark datasets, i.e., PPGDalia~\cite{reiss2019deep} and SPC2015~\cite{troika2014}.}
    \item \rev{A complete deployment of both individual TimePPG models and of the ActPPG framework on a low-power microcontroller (MCU), demonstrating our algorithms' suitability for the edge execution in terms of latency and energy consumption. }
\end{itemize}
The best performing TimePPG model, \textit{TimePPG-Big}, coupled with simple smoothing post-processing, achieves a Mean Absolute Error (MAE) of \unit[4.88]{BPM} on PPGDalia~\cite{reiss2019deep} (the largest public PPG dataset) and includes $\approx$ 232k trainable parameters. With an additional fine-tuning step, the MAE is further reduced to \unit[3.84]{BPM}.
At the other extreme, the smallest model in TimePPG, \textit{TimePPG-Small}, uses only 5k parameters while still reaching an acceptable MAE of \unit[5.63]{BPM}. 
\rev{When deployed on the STM32L4R9AII6, a popular low-power MCU by ST Microelectronics, \textit{TimePPG-Small} and \textit{TimePPG-Big} consume \unit[232]{$\mu$J}, and \unit[17.57]{mJ} per inference, with a latency of \unit[17.1]{ms}, and \unit[1289.5]{ms}, respectively. 
\textit{TimePPG-Small} outperforms previous deep learning solutions~\cite{reiss2019deep} on the PPG-Dalia dataset by 5.2-12000$\times$ and 3.8-4800$\times$ in terms of model size and the number of required operations respectively, while also improving the HR estimation accuracy (5.63 vs. 9.99/7.65 BPM of MAE).}
Thanks to the ActPPG adaptive approach, we obtain further Pareto points, reducing inference's computational complexity with a small accuracy degradation.
For instance, using a small random forest as movement detector and \textit{TimePPG-Small/Big} as HR estimators, we reduce the computation complexity by 50.3\% while losing only 0.39 BPM of MAE (for a benchmark in which 50\% of the samples have a ``high'' degree of MAs).
\rev{Deploying such configuration on the STM32L4R9AII6 MCU, we obtain as low as \unit[8.90]{mJ} of energy per inference.}

The rest of the paper is organized as follows.
Section~\ref{sec:background} provides the necessary background on PPG, on TCNs, and on our target hardware platform.
Section~\ref{sec:related} gives an overview of state-of-the-art HR estimation algorithms.
Section~\ref{sec:TimePPG} and Section~\ref{sec:ActPPG} describe our main contributions, i.e., TimePPG and ActPPG respectively.
Finally, Section~\ref{sec:results} presents the results and their discussion and Section~\ref{sec:conclusions} concludes the paper.

\section{Background}
\label{sec:background}
\subsection{Photoplethysmography}
Photoplethysmography (PPG) is a technique based on measuring the light absorption variations of blood vessels during the cardiac activity~\cite{tamura2014wearable}.
A PPG sensor consists of one or more Light-Emitting Diodes (LEDs) that continuously emit light to the skin and a photodetector (i.e., a photodiode) that measures variations of light intensity caused by blood flow, whose periodicity depends on the heart rate.
More specifically, the larger the blood volume variation, the greater the attenuation of the light emitted by the LED, resulting in a lower the current output on the photodiode. Therefore, a heartbeat can be associated with each peak in the PPG signal.
Indeed, many studies demonstrated that the second derivative of the PPG signal contains essential information for heart rate monitoring~\cite{troika2014}.

The simplicity of wearing a PPG sensor and the low-cost contribute to its increasing popularity as an alternative to ECG for HR monitoring~\cite{sviridova2015human}.
One of the major challenges in employing PPG signals is their significant dependency on the subject's movements, which negatively affect the measurement quality during daily activities, as first shown in~\cite{zhang2015combining}.
In particular, Motion Artifacts (MA) caused by the hands' movement alter the readings of the sensor and strongly impact the performance of the HR estimation, demanding ad-hoc algorithms for their remotion/reduction. As described in Section~\ref{sec:related}, the standard approach is to leverage additional intertial measurements (mostly, acceleration) to clean the PPG signal from MAs.
\rev{For more details on using PPG for HR estimation please refer to~\cite{PPG_survey}.}

\subsection{Temporal Convolutional Network}
\label{subsec:tcn} 
\begin{figure}[ht]
  \centering
  \includegraphics[width=0.4\columnwidth]{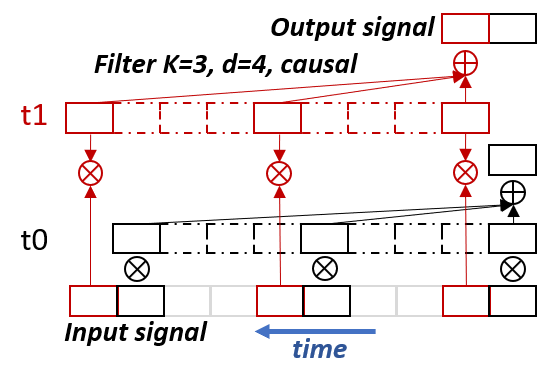}
  \vspace{-0.4cm}
  \caption{\rev{Generation of the first 2 output time samples in a TCN layer with $K$ = 3 and $d$ = 4.}}
  \label{fig:tcn_layer}
\end{figure}

\rev{In recent years, TCNs have achieved outstanding performance in many different time-series processing benchmarks, often resulting superior to Recurrent Neural Networks, which were previously considered the de-facto standard deep learning models for such tasks~\cite{bai2018empirical,zanghieri2019robust, 9283028}.}
TCNs are a sub-class of 1D-Convolutional Neural Networks (CNNs) esplicitly designed for the processing of time-series, whose peculiarity is in the use of \textit{causality} and \textit{dilation} in convolutional layers~\cite{bai2018empirical,lea2016temporal}.
\textit{Causality} constrains the convolution output $\mathbf{y}_{t}$ to depend only on inputs $\mathbf{x}_{\tilde{t}}$ with $\tilde{t} \leq t$, i.e., outputs are determined looking only to the ``past''. \textit{Dilation} is a fixed gap $d$ inserted between input samples processed by the filters that compose a convolutional layer. This has the benefit of increasing the receptive field of the convolution on the time axis, without requiring a larger number of parameters.
In summary, a convolutional layer in a TCN implements the following function:
\begin{equation}\label{eq:1d_conv}
\mathbf{y}_t^m = \sum_{i=0}^{K-1} \sum_{l=0}^{C_{in}-1} \mathbf{x}_{t-d\,i}^l \cdot \mathbf{W}_i^{l,m}
\end{equation}
where $\mathbf{x}$ and $\mathbf{y}$ are the input and output feature maps, $t$ and $m$ the output time-step and channel respectively, $\mathbf{W}$ the filter weights, $C_{in}$ the number of input channels, $d$ the dilation factor, and $K$ the filter size. \rev{Fig. \ref{fig:tcn_layer} shows a high-level scheme of the execution of a layer with $K$ = 3, and $d$ = 4.}
In the original paper~\cite{bai2018empirical}, TCNs were proposed as fully-convolutional architectures, but modern embodiments also include other common layers such as pooling and linear ones~\cite{ren2020cloud, zanghieri2019robust}, which have been used in our architecture exploration as well.

\subsection{Hardware platform}
\label{sec:hardware_setup}
\rev{Our work's main objective is to assess the feasibility of embedding deep learning-based HR estimation using PPG sensors and accelerometers in a wrist-worn wearable device. Therefore, rather than merely focusing on improving the HR estimation accuracy, our goal is to derive models that can be executed with low latency and energy consumption on a typical wearable platform, while retaining high accuracy.
Nowadays, most commercial wearables' digital ``brain'' consists of an ultra-low-power System-on-Chip, typically based on an ARM Cortex-M-class MCU.
Accordingly, we deploy all our models on the STM32L4R9I-EVAL evaluation board from STMicroelectronics, which features a Cortex-M4 core (the STM32L4R9AII6) with 640 kB of RAM, and 2MB of Flash.
The board has an average power consumption of 13.63 mW at 80 MHz \cite{st-L478}~\footnote{\rev{Notice that we use a platform with the same core but a larger memory space compared to the one adopted in our preliminary work of~\cite{Risso2021}, to enable the deployment of all TimePPG and ActPPG models.}}.}

\rev{
The selection of this platform is also driven by the availability of \emph{\textsc{Cube.AI}}, a software toolchain provided by STMicroelectronics to deploy neural networks on their MCUs. \emph{\textsc{Cube.AI}} allows the automatic conversion of pre-trained neural networks from high-level frameworks such as TensorFlow Lite and Keras into optimized C code for the execution of the network on the target.
Notably, at the time of submission of this article, \emph{\textsc{Cube.AI}} only allows the deployment of full-precision (\emph{float32}) dilated convolutions, while \emph{int8}-quantized convolutions with dilation are not supported.}
\section{Related Work}\label{sec:related}
\begin{table}[]
\centering
\footnotesize
\caption{State-of-the-art comparison table. Different MAE results correspond to the datasets in the Dataset column.}
\renewcommand{\arraystretch}{1.2}
\label{tab:SoA}
\begin{tabular}{|p{1.6cm}|c|c|c|c|c|p{1.4cm}|c|}
\hline
Work                                              & Dataset                                                                                             & Activities                                                                            & Sign.                                              & Pre-Processing                                                                                 & Algorithm                                                                                          & Post-Proc.                                                              & MAE                                                           \\ \hline \hline
\multicolumn{8}{|l|}{\textbf{Classical methods}}               \\ \hline\hline
TROIKA, 2014 \cite{troika2014}                & SPC2015$^*$                                                                                  & Rest, Running                                                                         & \begin{tabular}[c]{@{}c@{}}PPG, \\ Acc.\end{tabular} & \begin{tabular}[c]{@{}c@{}}0.5-4 Hz filtering, \\ Downsampling\end{tabular}                    & \begin{tabular}[c]{@{}c@{}}Signal decomp., \\ reconstruct., \\ spectral peak track\end{tabular} & \begin{tabular}[c]{@{}c@{}}th., \\ hist. track.\end{tabular}            & 2.34 BPM                                                      \\ \hline
JOSS, 2015 \cite{joss2015}                  & SPC2015$^*$                                                                                   & Rest, Running                                                                         & \begin{tabular}[c]{@{}c@{}}PPG, \\ Acc.\end{tabular} & \begin{tabular}[c]{@{}c@{}}0.5-4 Hz filtering, \\ Downsampling\end{tabular}                    & \begin{tabular}[c]{@{}c@{}}MMV, \\ spectral subtract\end{tabular}                               & \begin{tabular}[c]{@{}c@{}}th., \\ hist. track.\end{tabular}            & 1.28 BPM                                                      \\ \hline
SpaMa, 2016 \cite{spama2016}                 & \begin{tabular}[c]{@{}c@{}}SPC2015$^*$\\ SPC2015$^1$\\ Chon Lab$^2$\\ PPG-Dalia$^3$\end{tabular} & \begin{tabular}[c]{@{}c@{}}Rest, Running, \\ Rehab. ex.,\\ Rest, Running \\ 8 daily activities \end{tabular} & \begin{tabular}[c]{@{}c@{}}PPG, \\ Acc.\end{tabular} & \begin{tabular}[c]{@{}c@{}}0.5-3 Hz filtering, \\ Downsampling\end{tabular}                    & \begin{tabular}[c]{@{}c@{}}spectral filtering \\ based on PSD\end{tabular}                         & \begin{tabular}[c]{@{}c@{}}hist. track., \\ interpol.\end{tabular} & \begin{tabular}[c]{@{}c@{}}0.89 BPM\\ 3.36 BPM \\1.38 BPM \\ 11.06 BPM\end{tabular}   \\ \hline
Schack2017 \cite{schack2017computationally} & \begin{tabular}[c]{@{}c@{}} SPC2015$^*$\\ PPG-Dalia$^3$\end{tabular}                                                                                  & \begin{tabular}[c]{@{}c@{}}Rest, Running, \\ 8 daily activities \end{tabular}   & \begin{tabular}[c]{@{}c@{}}PPG, \\ Acc.\end{tabular} & \begin{tabular}[c]{@{}c@{}}0.5-6 Hz filtering, \\ Downsampling\end{tabular}                    & \begin{tabular}[c]{@{}c@{}}Corr.-based Freq. \\ indicating func., \\ FFT\end{tabular}              & th.                                                                     & \begin{tabular}[c]{@{}c@{}}1.32 BPM \\ 20.5 BPM\end{tabular}                                                     \\ \hline
FSM, 2018 \cite{chung2018finite}           & SPC2015$^1$                                                                                & \begin{tabular}[c]{@{}c@{}}Rest, Running, \\ Rehab. ex.\end{tabular}                  & \begin{tabular}[c]{@{}c@{}}PPG, \\ Acc.\end{tabular} & \begin{tabular}[c]{@{}c@{}}0.5-4 Hz filtering, \\ z-score scaling,\\ Downsampling\end{tabular} & Winer filtering                                                                                    & FSM                                                                     & 0.99 BPM                                                      \\ \hline
TAPIR, 2020 \cite{huang2020robust}           & \begin{tabular}[c]{@{}c@{}}SPC2015$^*$\\ SPC2015$^1$\\ PPG-Dalia$^3$\end{tabular} & \begin{tabular}[c]{@{}c@{}}Rest, Running, \\ Rehab. ex.,\\ 8 daily activities\end{tabular}                  & \begin{tabular}[c]{@{}c@{}}PPG, \\ Acc.\end{tabular} & \begin{tabular}[c]{@{}c@{}}0.5-4 Hz filtering\end{tabular} & \begin{tabular}[c]{@{}c@{}}Adaptive filter \\ Peak detection \\ Linear Transform.\end{tabular} & Notch filter                                                                     & \begin{tabular}[c]{@{}c@{}}2.5 BPM \\ 5.9 BPM \\ 4.6 BPM\end{tabular}    \\ \hline
\rev{Arunkmar, 2020 \cite{arunkumar2020robust}  }         & \begin{tabular}[c]{@{}c@{}}SPC2015$^*$\\ SPC2015$^1$\end{tabular} & \begin{tabular}[c]{@{}c@{}}Rest, Running, \\ Rehab. ex.\end{tabular}                  & \begin{tabular}[c]{@{}c@{}}PPG, \\ Acc.\end{tabular} & \begin{tabular}[c]{@{}c@{}}0.4-3.5 Hz filtering\end{tabular} & \begin{tabular}[c]{@{}c@{}}RLS, NLMS \\ MA reduction \\ FFT based HR track.\end{tabular} & Phase Voc.                                                                     & \begin{tabular}[c]{@{}c@{}}1.03 BPM \\ 1.89 BPM\end{tabular}   \\ \hline
CurToSS, 2020 \cite{zhou2020heart}           & \begin{tabular}[c]{@{}c@{}}SPC2015$^*$\\ SPC2015$^1$\\ PPG-Dalia$^3$\end{tabular} & \begin{tabular}[c]{@{}c@{}}Rest, Running, \\ Rehab. ex.,\\ 8 daily activities\end{tabular}                  & \begin{tabular}[c]{@{}c@{}}PPG, \\ Acc.\end{tabular} & \begin{tabular}[c]{@{}c@{}}0.5-4 Hz filtering\end{tabular} & \begin{tabular}[c]{@{}c@{}}SSR \\ Curve tracking\end{tabular}  & N/A                                                              & \begin{tabular}[c]{@{}c@{}}2.2 BPM \\ 4.5 BPM \\ 5.0 BPM\end{tabular}   \\ \hline \hline

\multicolumn{8}{|l|}{\textbf{Deep Learning}}      \\ \hline\hline
CNN, 2019 \cite{reiss2019deep}               & \begin{tabular}[c]{@{}c@{}}SPC2015$^*$\\ PPG-Dalia$^3$\end{tabular}                     & \begin{tabular}[c]{@{}c@{}}Rest, Running,\\ 8 daily activities\end{tabular}           & \begin{tabular}[c]{@{}c@{}}PPG, \\ Acc.\end{tabular} & \begin{tabular}[c]{@{}c@{}}STFT,\\ 0-4 Hz filtering\end{tabular}                               & CNN                                                                                                & N/A                                                                     & \begin{tabular}[c]{@{}c@{}}4 BPM \\ 7.65 BPM\end{tabular}     \\ \hline
\rev{PPGNet, 2019 \cite{shyam2019ppgnet}  }& \begin{tabular}[c]{@{}c@{}}SPC2015$^*$\\ SPC2015$^1$\end{tabular}                       & \begin{tabular}[c]{@{}c@{}}Rest, Running, \\ Rehab. ex.\end{tabular}                  & \begin{tabular}[c]{@{}c@{}}PPG\end{tabular}  & \begin{tabular}[c]{@{}c@{}}0.4-18 Hz filtering,\\ z-score scaling\end{tabular}                 & Inception+LSTM                                                                                           & N/A                                                                     & \begin{tabular}[c]{@{}c@{}}3.36 BPM  \\ 12.48 BPM\end{tabular} \\ \hline
CorNet, 2019 \cite{cornet2019}                & \begin{tabular}[c]{@{}c@{}}SPC2015$^*$\\ SPC2015$^1$\end{tabular}                       & \begin{tabular}[c]{@{}c@{}}Rest, Running, \\ Rehab. ex.\end{tabular}                  & \begin{tabular}[c]{@{}c@{}}PPG\end{tabular}  & \begin{tabular}[c]{@{}c@{}}0.4-18 Hz filtering,\\ z-score scaling\end{tabular}                 & CNN+LSTM                                                                                           & N/A                                                                     & \begin{tabular}[c]{@{}c@{}}4.67 BPM  \\ 5.55 BPM\end{tabular} \\ \hline
\rev{BinCorNet, 2020 \cite{rocha2020binary}}                & \begin{tabular}[c]{@{}c@{}}SPC2015$^*$\\ SPC2015$^1$\end{tabular}                       & \begin{tabular}[c]{@{}c@{}}Rest, Running, \\ Rehab. ex.\end{tabular}                  & \begin{tabular}[c]{@{}c@{}}PPG\end{tabular}  & \begin{tabular}[c]{@{}c@{}}0.4-18 Hz filtering,\\ z-score scaling\end{tabular}                 & Bin. CNN+LSTM                                                                                           & N/A                                                                     & \begin{tabular}[c]{@{}c@{}}6.78 BPM  \\ 7.32 BPM\end{tabular} \\ \hline
\rev{Chung, 2020 \cite{chung2020deep}}                & \begin{tabular}[c]{@{}c@{}}SPC2015$^*$\end{tabular}                       & \begin{tabular}[c]{@{}c@{}}Rest, Running.\end{tabular}                  & \begin{tabular}[c]{@{}c@{}}PPG, \\ Acc.\end{tabular}  & \begin{tabular}[c]{@{}c@{}}0.4-18 Hz filtering,\\ z-score scaling\end{tabular}                 & CNN+LSTM                                                                                           & N/A                                                                     & \begin{tabular}[c]{@{}c@{}}1.46 BPM\end{tabular} \\\hline
DeepHeart, 2021 \cite{deepheart2021}               &\minor{SPC2015$^*$}   &\minor{Rest, Running.}                & \begin{tabular}[c]{@{}c@{}}\minor{PPG,} \\ \minor{Acc.}\end{tabular}  & \minor{0.4-5 Hz filtering}             & \minor{DnCNN + spectrum analysis }                                                                                          & \begin{tabular}[c]{@{}c@{}}\minor{err. check,} \\ \minor{calibration.}\end{tabular}                                                                    & \minor{1.61 BPM} \\ \hline

Our Work                                          & \begin{tabular}[c]{@{}c@{}} PPG-Dalia$^3$\end{tabular}                     & \begin{tabular}[c]{@{}c@{}} 8 daily activities\end{tabular}           & \begin{tabular}[c]{@{}c@{}}PPG, \\ Acc.\end{tabular} & \begin{tabular}[c]{@{}c@{}}0.5-4 Hz filtering\end{tabular}                               & \begin{tabular}[c]{@{}c@{}}TCNBest\end{tabular}  & \begin{tabular}[c]{@{}c@{}}th, \\ finetuning\end{tabular}                                                                  & \begin{tabular}[c]{@{}c@{}}3.84 BPM \end{tabular}                                        \\ \hline
\multicolumn{8}{l}{$^*$ 12 subjects $^1$ 23 subjects $^2$ 10 subjects $^3$ 15 subjects  }      \\

\end{tabular}
\end{table}
Heart rate monitoring through wrist-worn PPG sensors is a relatively new application, attracting  significant interest in industry and academia. 
The main challenges that this approach has to tackle are related to \textit{i)} the accuracy, often measured as the $MAE =\mathbb{E}(|BPM_{true}-BPM_{prediction}|)$ and \textit{ii)} the tight power and energy constraints that have to be met in order to execute locally on battery-operated wrist-worn devices;
in fact, these algorithms are usually executed on a smartwatch, with a MCU operating at frequencies in the range of 10s of MHz and having a power envelope lower than 100 mW.

The first algorithms for PPG-based HR estimation were based on simple peak tracking methods. Among them, a representative example is found in~\cite{RollingMean}, where the authors propose an improved peak detection algorithm called Adaptive Threshold (AT), which removes false peaks using an adaptive refractory period.

More recent algorithms can be split into two main categories: classical model-driven approaches, which are mostly based on adaptive-filtering combined with peak tracking, and data-driven ones, based on machine/deep learning. The leading solutions proposed in the literature are summarized in Table~\ref{tab:SoA}.

In the model-driven category, the seminal work of~\cite{troika2014} paved the way to the algorithmic exploration in this field, proposing a new three-stage algorithm based on signal decomposition, spectrum estimation, and spectral peak tracking called TROIKA. TROIKA has been tested on the public dataset called SPC Cup 2015 (SPC2015), released together with the paper and comprising 12 subjects, achieving a MAE of 2.34 BPM.
The same authors have also proposed an improvement of this algorithm in~\cite{joss2015}, where a spectral difference with the acceleration spectrum is used to clean the PPG spectrum from MAs, reducing the MAE on the same dataset to 1.28 BPM.
Following the same MA cleaning approach, in~\cite{mashhadi2015heart}, the authors propose to suppress them using a Singular Value Decomposition (SVD) applied on the acceleration data to extrapolate periodic artifacts. Coupled with an Iterative Method with Adaptive Thresholding (IMAT), this solution reduces the MAE on SPC2015 to 1.25 BPM.
\rev{Similar MA reductions are also used in~\cite{arunkumar2020robust,schack2017computationally} to clean the PPG signal, and a FFT to track the HR, achieving 1.32 and 1.03 BPM of MAE on SPC2015, whereas the authors of~\cite{temko2017accurate, chung2018finite}} use Wiener filtering, further improving the performance on SPC2015 to 0.99 BPM of MAE.
Lastly, the best performing approach on this dataset is SpaMa, proposed in~\cite{spama2016} and based on a complex and computationally intensive five-step pipeline. This algorithm reduces the MAE to just  0.89 BPM.

More recently, Reiss et al. \cite{reiss2019deep} released a new PPG-based HR tracking dataset, significantly larger than SPC2015. The dataset, called PPG-Dalia, includes 15 subjects monitored while performing eight different daily activities.
In~\cite{zhou2020heart, huang2020robust} the authors proposed two new model-driven HR estimation solutions, tested both on SPC2015 and PPG-Dalia. The first, CurToSS~\cite{zhou2020heart}, exploits sparse signal reconstruction and tracking of the curve in both acceleration and PPG signals, obtaining 2.2 BPM of MAE on SPC2015. The second, TAPIR~\cite{huang2020robust}, performs linear transformations in the time domain, strongly reducing the computational complexity and achieving a MAE of 2.5 BPM on the same dataset.

In general, all model-driven algorithms include many free parameters, leading to a high risk of overfitting.
For instance, when two of the best-performing algorithms on the SPC2015 dataset are tested on the more extended PPG-Dalia, they obtain dramatically worse MAEs, from 0.89 to 11.06 BPM for~\cite{spama2016}, and from 1.32 to 20.5 BPM for~\cite{schack2017computationally}.
Similarly, the only two model-driven algorithms directly developed for PPG-Dalia~\cite{zhou2020heart, huang2020robust} obtain the best performance on that dataset, 5.0 and 4.6 BPM, respectively, but achieve lower accuracy than the state-of-the-art for SPC2015.
Moreover, adaptive filters, which are the core of nearly all the algorithms analyzed so far, are computationally intensive and not well suited for a real-time embedded application. Indeed, to the best of our knowledge, none of the aforementioned model-driven algorithms has been ported to a MCU-class platform such as those found on wrist-worn devices in order to evaluate its execution latency and energy consumption.

In the last years, motivated by the increasing success of machine/deep learning approaches in several other bio-signal processing applications, such as gesture recognition~\cite{zanghieri2019robust}, seizure detection~\cite{burrello2020hyperdimensional, burrello2020ensemble} and brain-computer interfaces~\cite{zhang2019survey}, researchers have started to explore solutions based on Convolutional Neural Networks (CNN) and Recurrent Neural Networks (RNN) for HR tracking.
In \cite{reiss2019deep}, in conjunction with the publication of PPG-Dalia,  the authors presented a CNN architecture coupled with a short-time Fourier transform, which outperforms two of the best classical methods~\cite{spama2016,schack2017computationally} on the new dataset.
\minor{Furthermore, the works of \cite{cornet2019, shyam2019ppgnet,chung2020deep, deepheart2021} achieve comparable results to the classical methods, using either pure deep learning architectures (e.g., CNN+LSTM) or denoising CNNs (DnCNNs) to remove motion artifacts followed by spectrum analysis applied to the clean PPG signals.}

Deep learning solutions are known to achieve better generalization than classical ones on many tasks, but their application in this domain presents several challenges. Indeed, these models typically have many learned parameters (i.e., large memory footprints) and require a large number of operations for inference. Therefore, their deployment on memory-constrained MCUs, with low energy consumption and respecting real-time latency constraints, is not trivial.
\rev{A first attempt to deploy a deep neural network for HR tracking has been made in BinaryCorNET \cite{rocha2020binary}, where the authors binarized the network of \cite{cornet2019} and deployed it on dedicated hardware implemented in both ASIC and FPGA technologies, achieving just \unit[56.1]{$\mu$J} of energy consumption per classification, with a slightly higher MAE of 6.78 BPM on the SPC2015 dataset.
To the best of our knowledge, we are the first to investigate the embedding of deep learning models for PPG-based HR tracking onto programmable, general purpose edge MCUs rather than custom hardware. Furthermore, we are also the first to use TCNs (described in Section~\ref{subsec:tcn}) for this task.
We explicitly focus on reducing the complexity and energy consumption of these models, exploring efficient architectures as well as an adaptive HR tracking solution that combines multiple models to improve efficiency (the \textit{ActPPG} framework), while maintaining a low error.}

\section{TimePPG: Temporal Convolutional Networks for HR estimation}
\label{sec:TimePPG}
This section describes the exploration of different Temporal Convolutional Networks (TCN) architectures for HR tracking in the accuracy vs. complexity space, aimed at finding accurate and light-weight solutions to be deployed on the hardware described in Section~\ref{sec:hardware_setup}.
We restrict the search space to architectures that can be executed on the target MCU, the STM32L4, while respecting a real-time constraint of 2s per inference. This constraint is in accordance with previous work~\cite{reiss2019deep, huang2020robust}, and is equal to the time-shift (slide) between two consecutive input samples in the PPG-Dalia dataset (see Section~\ref{sect:arch_opt_seed}).

\begin{figure}[t]
 \centering
\includegraphics[width=0.6\columnwidth]{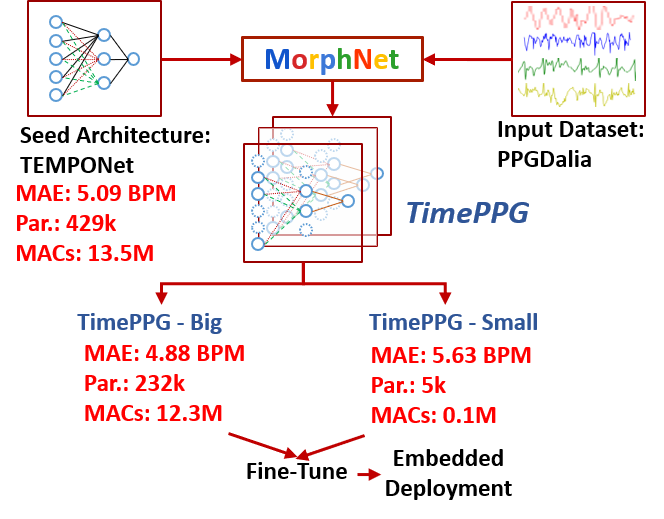}
  \caption{Proposed NAS and deployment flow. \minor{The original TCN (seed architecture) and the input dataset are fed to the MorphNet NAS tool~\cite{gordon2018morphnet} to generate a set of optimized models called TimePPG.} In red, the architectural parameters and MAE of the seed architecture and of two TimePPG Pareto points.}
  \label{fig:flow}
\end{figure}
Fig.~\ref{fig:flow} depicts the complete flow that we use to generate a set of TCNs.
As shown, we explore the design space through a Neural Architecture Search (NAS) tool called MorphNet \cite{gordon2018morphnet}, which we enhanced to work on 1D dilated convolutions instead of its original target, i.e., the 2D layers found in standard CNNs for computer vision. 
\minor{MorphNet receives as input the training dataset, together with an original TCN, called ``seed'', which serves as the starting point for the architectural exploration. The NAS then optimizes the seed in different ways, generating a set of optimized models, offering different trade-offs in terms of HR tracking error versus complexity, which we call TimePPG. The MAE obtained by the seed on the PPG-Dalia dataset is reported in Fig.~\ref{fig:flow}, together with its complexity in terms of the number of Parameters (Par.) and MAC operations.  The same data are also reported for two extreme models in the TimePPG set, i.e., the most accurate network (TimePPG- Big) and the smallest one (TimePPG-Small), to show the positive effect of the NAS application. These results are analyzed in details in Section~\ref{sec:results}. TimePPG models, optionally fine-tuned on the patient under test, can then be deployed onto a wearable device, choosing the TCN version that best matches the constraints of the target hardware.}
\minor{We describe our design space exploration process in Section~\ref{sect:arch_opt_mn}, while Section~\ref{sect:arch_opt_seed} describes the seed network} and Section~\ref{sect:arch_opt_improv} discussed two additional steps to improve the accuracy of our models.

\subsection{Design Space Exploration} 
\label{sect:arch_opt_mn}
NAS tools automatically generate novel NN architectures for a given task by acting on hyper-parameters such as the depth and the width of the network, the type of layers included, the connections between layers, etc~\cite{tan2019mnasnet, lin2020mcunet, cai2018proxylessnas}.
While NAS is receiving increasing attention, standard tools mostly target complex computer vision tasks, leading to large and computationally-intensive NNs and requiring an enormous number of training iterations. 
Therefore, in this work, we decided to apply MorphNet \cite{gordon2018morphnet} a tool that belongs to a category of recently developed light-weight NAS approaches called \textit{DmaskingNAS}, whose complexity is comparable to that of a \textit{single} NN training, at the price of a smaller search space. To obtain this result, DmaskingNAS tools generate their outputs as modified versions of a single \textit{seed network}.

In particular, MorphNet limits the optimization to the \textit{number of channels} (i.e., feature maps) in each layer, keeping the rest of the topology identical to the one of the seed. The number of channels is tuned to reduce the total memory footprint and number of required Multiply-and-Accumulate (MAC) operations of the network while retaining as much as possible the accuracy of the seed.
This is obtained thanks to a two-step algorithm, which we extended to work with TCN layers. In the first \textit{pruning} step, the network size is reduced by applying group Lasso~\cite{yuan2006model}, a training regularizer that forces the weights relative to entire channels (rather than single weights) to small magnitudes. After this regularization, channels whose total magnitude is inferior to a tunable threshold are eliminated.
The second step is an \textit{expansion}, in which the number of channels of all layers is up-scaled by a constant factor, called width multiplier. The goal is to recover part of the obvious performance penalty resulting from the first step. 
Noteworthy, as mentioned above, the complexity of MorphNet training is only slightly higher than that required to train the seed architecture.
For more details on this tool, we refer the reader to the original paper~\cite{gordon2018morphnet}.

Besides the seed network, MorphNet requires two other input parameters, i.e., the \textit{pruning strength, $p_t$}, and the \textit{pruning threshold} $th_{p}$, which control the strength for the group Lasso regularization and the number of eliminated channels in the pruning step.
In order to explore the design space, we swept these two parameters between $10^{-6}$ and $10^{-3}$ ($p_t$) and $10^{-2}$ to $10^{-1}$ ($th_p$), further filtering all outputs that resulted in a computation time $>$ 2 seconds. The results of this exploration are the different TCNs that compose TimePPG.

\subsection{Seed Network}\label{sect:arch_opt_seed}
As a seed network for our exploration, we used TEMPONet~\cite{zanghieri2019robust} a TCN which shows impressive results on another bio-signal processing task, i.e., EMG-based gesture recognition.
The TEMPONet architecture can be split into a modular \textit{feature extractor}, composed of 3 convolutional blocks, and a \textit{classifier}, composed of three linear layers.
In turn, each convolutional block is formed by two dilated convolutions to increase the receptive field, a strided convolution, and an average pooling layer to shrink the time dimension.
The output channels increase in each successive block to 32, 64, and 128, respectively. 
All layers use ReLU activations and batch normalization~\cite{ioffe2015batch}.
Further details on our seed network can be found in~\cite{zanghieri2019robust}.

In order to adapt TEMPONet to our task, we changed the first layer to match the dimensions of each sample in the PPG-Dalia dataset. In particular,
the network takes as input raw sensor data generated by the PPG-sensor and by a tri-axial accelerometer.
The data provided to the algorithm are sampled at \unit[32]{Hz} and fed to the model in sliding windows of \unit[8]{s}, with a slide of 2s. Therefore, each input sample has four channels, ($Ch_{in}=4$), and 256 time-steps ($t\in [0:255]$).
We also replaced the final classification layer of TEMPONet with a single neuron in order to adapt it to a scalar regression task and we changed the loss function used for training from Categorical Cross Entropy to LogCosh. We found that, as in~\cite{neuneier1998train}, LogCosh outperforms both RMSE and MAE losses, favoring the convergence near the minimum, thanks to its smoother behavior in that point.


\subsection{Performance improvements}\label{sect:arch_opt_improv}

\subsubsection{Post-Processing}
\label{sect:arch_opt_pp}
Orthogonally to our design space exploration, we applied a low-complexity smoothing post-processing to the outputs of all models in TimePPG, to further improve their accuracy. This post-processing is motivated by the fact that, despite being more accurate on average than model-driven approaches, data-driven ones such as deep learning algorithms may sometimes incur huge errors, especially when the processed inputs differ from those seen in the training phase.
Fortunately, in this particular task, these errors can be easily filtered by considering the compatibility of TCN estimations with human physiology, and in particular with the dynamics of the heart rate. Specifically, we impose a limit on the maximum variation of the estimated HR over time.
To this end, we compare the latest TCN prediction with the mean of the previous N predictions, $E_{HR}[N]$: if the difference is larger than a threshold $P_{th}$, the HR is \textit{clipped} to $E_{HR}[N]$ $\pm$ $P_{th}$.
In this work, we set N to 10 and $P_{th}$ = $\nicefrac{E_{HR}[N]}{10}$, identical for all patients. \rev{When deployed on the target MCU, we found that this post-processing has negligible time and space complexity even with respect to the smallest HR estimation models.}

\subsubsection{Fine-Tuning} 
\label{sect:arch_opt_ft}
\begin{figure}[t]
  \centering
\includegraphics[width=0.7\columnwidth]{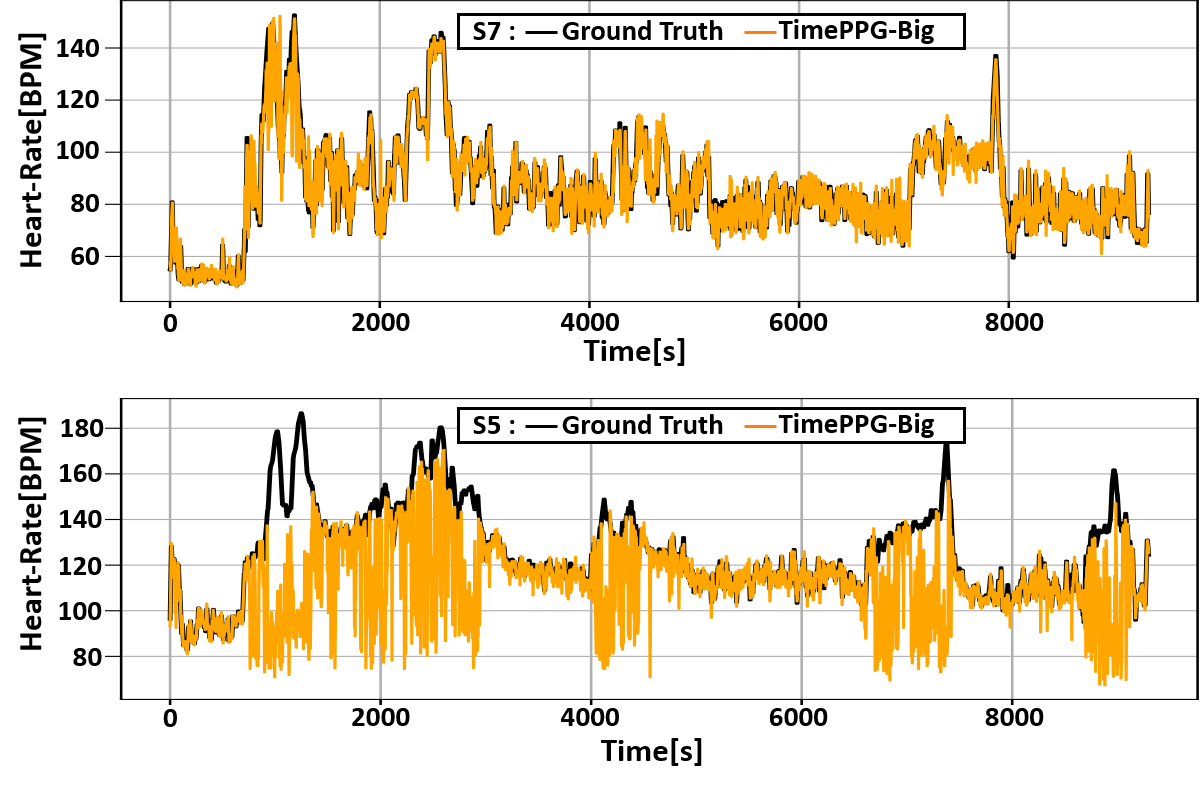}
  \caption{Ground truth and prediction of a well tracked patient (S7) and of the worst one (S5). Values above 140 BPM are not well estimated by our algorithm.}
  \label{fig:trace}
\end{figure}

In one of our experiments of Section~\ref{sec:results} we have also analyzed the performance of our models after an additional fine-tuning step.
This is motivated by the fact that the lack of large datasets for PPG-based HR monitoring (public ones are limited to $<20$ subjects) can hinder the performance of deep learning solutions, which notoriously benefit from large amounts of training data.
In particular, input data never seen during training are challenging to predict.
\minor{Therefore, subjects with \textit{very large} HR values, that are rare in the rest of the dataset, are poorly tracked by our TCNs and, more generally, by data-driven algorithms. Very low HRs also have the same problem, but since they are typically associated to static activities with fewer MAs, the tracking problem is inherently easier for them.}
To underline this effect, Fig. \ref{fig:trace} shows the accurate tracking obtained using the (not fine-tuned) TimePPG-Big outputs on a subject whose HR is in the interval $[50,140]$ BPM (S7), and the much less accurate results obtained on S5, when the HR $>$ is 140 BPM, i.e., in the right tail of PPG-Dalia data distribution.
We claim that this phenomenon could disappear in the presence of a more extended dataset and is not an intrinsic limitation of TimePPG.
\minor{To demonstrate it, we \textit{simulate} the effect of a larger dataset, which would include abundant samples of the entire range of possible HR values that our models are expected to predict, by means of fine-tuning. Specifically,}
we fine-tuned our networks on the first portion  (25\%) of each subject's data before testing on the remaining portion, obtaining a substantial performance improvement on patients whose HR is weakly represented in the training set.
%
%
Note that, while this step is hardly reproducible in the field since fine-tuning would require collecting ground truth HR data from the user, e.g., through an ECG,
\minor{we apply it only to mimic the results that could be obtained by training our models on a larger dataset, including data similar to all test subjects.}
\section{ActPPG: Rest-Movement based HR estimation }
\label{sec:ActPPG}
In this section we detail the \textit{ActPPG} framework, specifically designed to adaptively combine multiple HR tracking models in a single prediction solution, using movement to assess the difficulty of the HR estimation. 
In Section~\ref{sect:act_motiv} we provide the motivation and the intuition behind our framework, showing an example of its functionality by combining the AT model of \cite{RollingMean} and the smallest TimePPG model, hereafter called TimePPG-Small.
Then, we detail the different components of ActPPG in Section~\ref{sect:act_modules}.
\subsection{Motivation}\label{sect:act_motiv}
Our framework aims to reduce the energy consumption of on-chip HR monitoring by combining the predictions of two different classifiers at runtime, one more accurate but more energy-hungry, and the other less accurate but more energy-efficient. 

Inspired by big-little neural networks~\cite{park2015big,JahierPagliari2018a,Mocerino2020,Daghero2020}, we aim at employing the heavier classifier only for those inputs for which the lighter one would fail. Achieving this goal would allow maintaining the same performance of the larger and more accurate classifier while decreasing the average energy consumption per input, depending on the frequency of ``difficult'' samples.
Standard big-little models try to achieve this goal in an application-agnostic way, by sequentially executing first the little model and then optionally the big one, when the former's output has a low \textit{confidence score} (see~\cite{park2015big,JahierPagliari2018a,Mocerino2020,Daghero2020} for details). Therefore, for ``difficult'' inputs, these approaches require executing \textit{two models} and incur an energy overhead.
In contrast, we propose an application-specific adaptive system that always requires a \textit{single inference} \rev{on PPG data} per input. Specifically,
we use the amount of movement as a proxy of the HR prediction difficulty and use it to discriminate between easy and difficult samples.
Our framework relies on two hypotheses:

\textbf{1) MAE $\propto$ Mov.} Regardless of the model used, as the amount of movement of a subject increases, the HR prediction becomes increasingly less accurate because of MAs.

\textbf{2) Modelling Difficulty.}
The performance difference between a simple and small HR predictor and a larger and more complex one is mostly due to the latter's ability to obtain more accurate predictions for inputs affected by high movement.
In other words, taken two models $M1$ and $M2$, with MAE$_{M1}$ $>$ MAE$_{M2}$, we observe that often MAE$_{M1}^{rest}$ $\sim$ MAE$_{M2}^{rest}$ for rest windows while MAE$_{M1}^{mov}$ $>>$ MAE$_{M2}^{mov}$ for high movement windows.

\begin{figure}[ht]
  \centering
  \includegraphics[width=0.6\columnwidth]{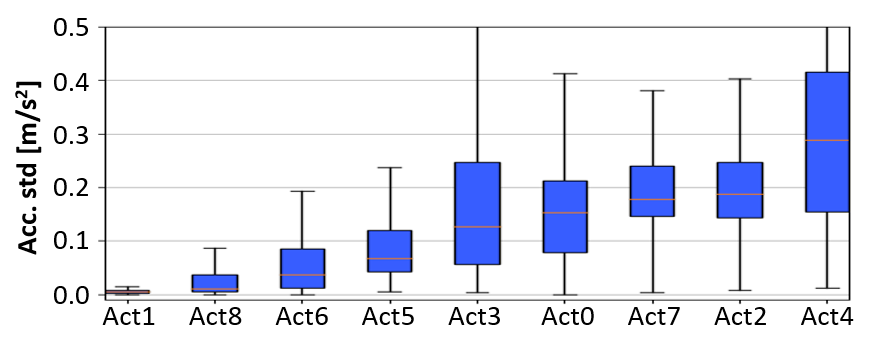}
  \caption{PPG-Dalia activities ordered by increasing standard deviation of the accelerometer signal.}
  \label{fig:acc_order}
  \end{figure}

\begin{figure}[ht]
  \centering
  \includegraphics[width=0.6\columnwidth]{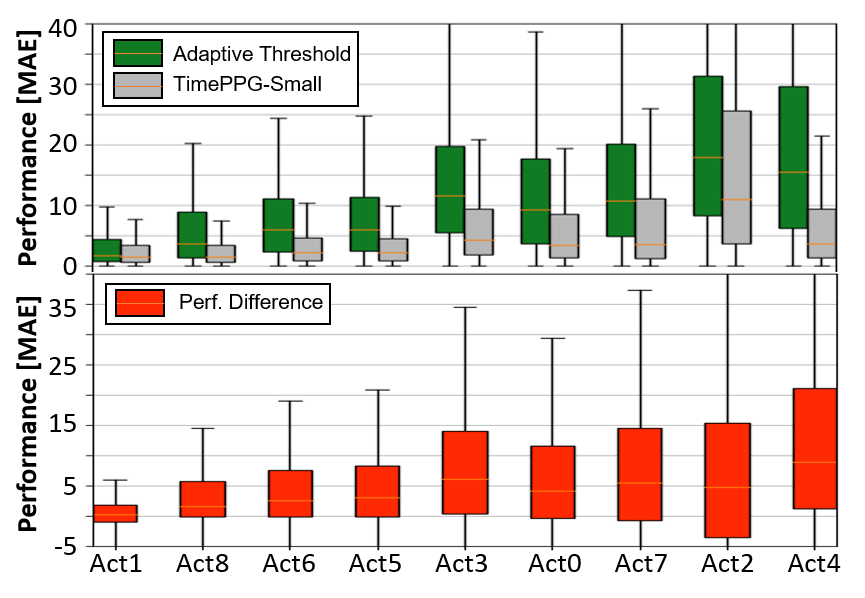}
  \caption{Performance of two HR tracking algorithms (Adaptive Threshold, TimePPG-Small) on each PPG-Dalia activity (top). Performance difference between the same two algorithms (bottom).} \label{fig:Act_order}
  \end{figure}

To demonstrate these intuitions, we propose an example comparing two algorithms of different complexity and accuracy: the Adaptive-Threshold (AT) described in \cite{RollingMean} and the TimePPG-Small network, i.e., the most energy-efficient solution in TimePPG collection.
First, we grouped samples relative to each of the 8 daily activities found in the PPG-Dalia dataset and sorted them by increasing movement, as shown in Figure~\ref{fig:acc_order}.
\rev{As an example, the sorting has been done using as a measure of the ``amount of movement'' the increasing average standard deviation (std) on the 3 axes of the accelerometer signal in each 8s window of the dataset.}
%
%
Then, we tested the two algorithms on each activity separately.
Figure~\ref{fig:Act_order} reports the median (orange horizontal segments) and first/third quantiles of the algorithms' performance over all the different activities.

Coherently with our first hypothesis, the performance of both algorithms deteriorates for activities with higher movements, going from a minimum of 1.69 [0.70, 4.35] / 1.45 [0.60, 3.45] BPM (AT/TimePPG-Small medians with $[$first quantile, third quantile$]$) for activity 1, i.e., \textit{sitting}, to 17.88 [8.27, 31.31] / 11.0 [3.61, 25.60] BPM for activity 2, \textit{climbing the stairs}, which has the second-largest acceleration std. 


\begin{figure}[ht]
  \centering
  \includegraphics[width=1\columnwidth]{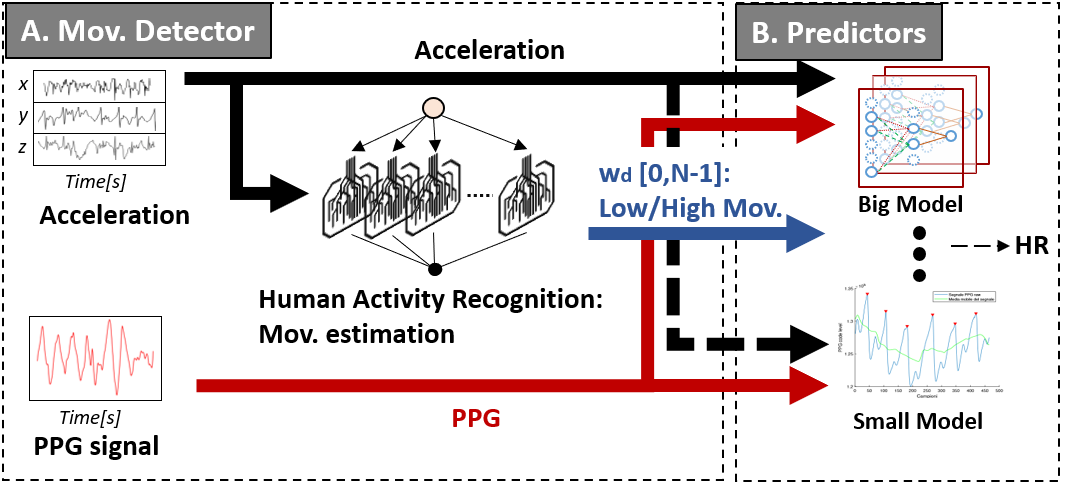}
  \caption{Flowchart of the ActPPG framework. \minor{On the left, a human activity recognition (HAR) model (a random forest in our implementation) estimates the amount of MAs in the current input based on acceleration data. On the right, one of the N available heart rate tracking models is invoked based on the HAR result, with ``bigger'' models selected only for high-movement windows. In turn, the HR tracking model may use PPG data only or both PPG and acceleration.}}
  \label{fig:ActPPGflow}
  \vspace{-0.2cm}
  \end{figure}

Moreover, in accordance with our second hypothesis, the performance difference between the two algorithms also increases for high movement activities, as shown in the bottom of Figure \ref{fig:Act_order}. Specifically, it goes from a minimum difference of 0.23 [-0.95, 1.81] BPM for sitting to a maximum of 8.87 [1.26, 21.0] BPM for \textit{cycling}, the activity with the highest acceleration std.
Note that these assumptions also hold for different algorithms, e.g., for different TimePPG models, as demonstrated in the following sections.
\subsection{ActPPG Modules}\label{sect:act_modules}
\rev{The ActPPG framework is composed of two main modules, as shown in Figure~\ref{fig:ActPPGflow}.
The framework is fed with two signals, the PPG and the 3-axial acceleration. 
PPG data are the main inputs employed to predict HR in wrist-worn devices and are thus employed by all algorithms.
In this work, we process a new window every 2s, and we use 8s as window length, following the setup of~\cite{reiss2019deep}.}
On the other hand, 3D acceleration does not contain essential features for HR but is used by many state-of-the-art algorithms to ``clean'' the PPG signal from the MAs caused by the subjects' hand movement, as detailed in Section~\ref{sec:related}.
In our framework, we use acceleration data for two different tasks. First, we process them to determine whether a window is considered a ``rest'', i.e., a window with low MAs, or a ``movement'', with a high probability of  MAs corruption. Second, the acceleration signal could also be (optionally) employed by the HR predictors directly to supplement the PPG signal for MA rejection~\cite{troika2014, reiss2019deep}. 

\rev{The core component of ActPPG is the \textit{movement detector} module (Figure~\ref{fig:ActPPGflow}-A), whose goal is to assess the difficulty of the HR estimation for a given window.
In particular, based on the assumptions of Section~\ref{sect:act_motiv}, we design a detector that uses the acceleration as input and gives a binary 0/1 output to distinguish between low and high movement windows.
Note that this approach can be easily generalized to predict an ordinal output on a scale [0, N-1], where N-1 represents windows with the highest movement. One of N ordered HR predictors (from the smallest and less accurate to the largest and most accurate) can then be selected based on the window difficulty. In this work, we limit our exploration to the case of N=2.
As detector, we use a lightweight random forest (RF) composed of 8 trees with a maxium depth of 5.
We feed the RF with mean value, standard deviation, energy, and number of peaks of the acceleration signal. Combining these features with a RF allows to improve the final HR tracking performance, with respect to distinguishing between low and high movement with a simple threshold on a single feature (e.g., the acceleration std), and has a negligible impact on the overall inference complexity.}

The \textit{predictors} module (Figure~\ref{fig:ActPPGflow}-B) performs the  actual HR prediction.
This block receives as input the \emph{window difficulty}, $w_d$, (0 to N-1 ordinal variable, with N=2 in our work), the PPG data, and optionally the 3D acceleration.
The $w_d$ variable is used as a switch to select the predictor to use for a given window.
The predictors are sorted by performance offline, and more performant algorithms are associated with higher values of $w_d$.
Noteworthy, all predictors are selected on the Pareto frontier in the accuracy vs. complexity plane.
Therefore, when ordering by performance, we automatically associate more lightweight models to lower values of $w_d$.


\section{Experimental Results}
\label{sec:results}
\minor{In this section, we first describe in detail our target datasets.}
Then, we assess the performance of the individual HR estimation models from the TimePPG family and of the movement detector.
%
%
Finally, we show how the complete ActPPG framework can be used to effectively trade-off the accuracy of HR tracking with the inference complexity. 
We then assess the energy consumption, memory footprint, and latency of our solutions when deployed on the platform described in Section~\ref{sec:hardware_setup}. 
Finally, we compare the individual TimePPG models with \rev{recently developed techniques from the state-of-the-art. Since some of these were not tested on PPG-Dalia, for this comparison we also report results on the SPC2015 dataset}.
Software experiments are performed using Python 3.6 and TensorFlow 1.14 \cite{tensorflow2015}.

\subsection{\minor{Target Datasets}}
We evaluate our models mainly on the \textit{PPG dataset
for motion compensation and heart rate estimation in Daily-Life Activities} (PPG-Dalia)~\cite{reiss2019deep}. At the time of writing, PPG-Dalia is, to the best of our knowledge, the largest publicly available dataset for PPG-based HR estimation. 
It includes PPG data, 3D acceleration, reference ECG, HR and activity labels, computed on a moving window of 8 seconds with 6 seconds overlap. A total of 37.5 hours have been monitored from 15 subjects, eight female and seven male, with age in the interval 21-55.
Two commercial devices are used to collect data: the RespiBAN \cite{respiban} for the reference ECG, and the wrist-worn Empatica E4 \cite{empatica} for PPG and acceleration data.
\minor{For further details on PPG-Dalia, refer to~\cite{reiss2019deep}.
In Section~\ref{sec:soa_compare}, we also evaluate some of our TimePPG models on the smaller SPC Cup 2015 (SPC2015) dataset~\cite{troika2014}.
For both datasets, we follow the leave-one-subject-out validation schemes proposed in~\cite{reiss2019deep} and \cite{troika2014} respectively. Accordingly, all results reported in the rest of this section refer to the test subjects which, except for the case of fine-tuning on PPG-Dalia (see Section~\ref{sec:soa_compare}) are completely unseen during training.}
The state-of-the-art results have been taken directly from the original papers when available.

\subsection{TimePPG Results}
\label{sec:timeppg}
\begin{figure}[ht]
  \centering
  \includegraphics[width=0.6\columnwidth]{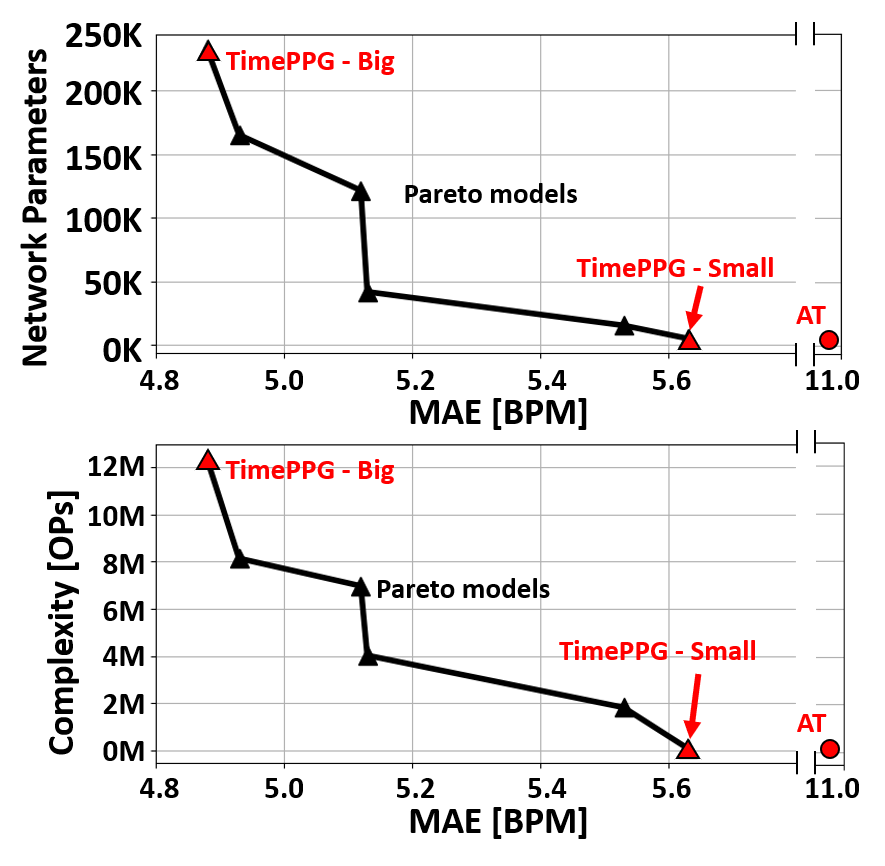}
  \caption{TimePPG and AT Pareto models \rev{on the PPG-Dalia dataset} in the MAE vs. Parameters space and in the MAE vs. complexity (number of OPs) space.}
  \label{fig:single_models}
  \end{figure}

Figure~\ref{fig:single_models} depicts some of the Pareto-optimal models extracted from the TimePPG family (black triangles) in the MAE vs. model size and the MAE vs. complexity (i.e., number of arithmetic operations -- OPs) spaces on the PPG-Dalia dataset.
These results refer to the algorithms' output after the smoothing post-processing step described in Section~\ref{sect:arch_opt_improv}, without fine-tuning.
\begin{figure}[ht]
  \centering
  \includegraphics[width=0.75\columnwidth]{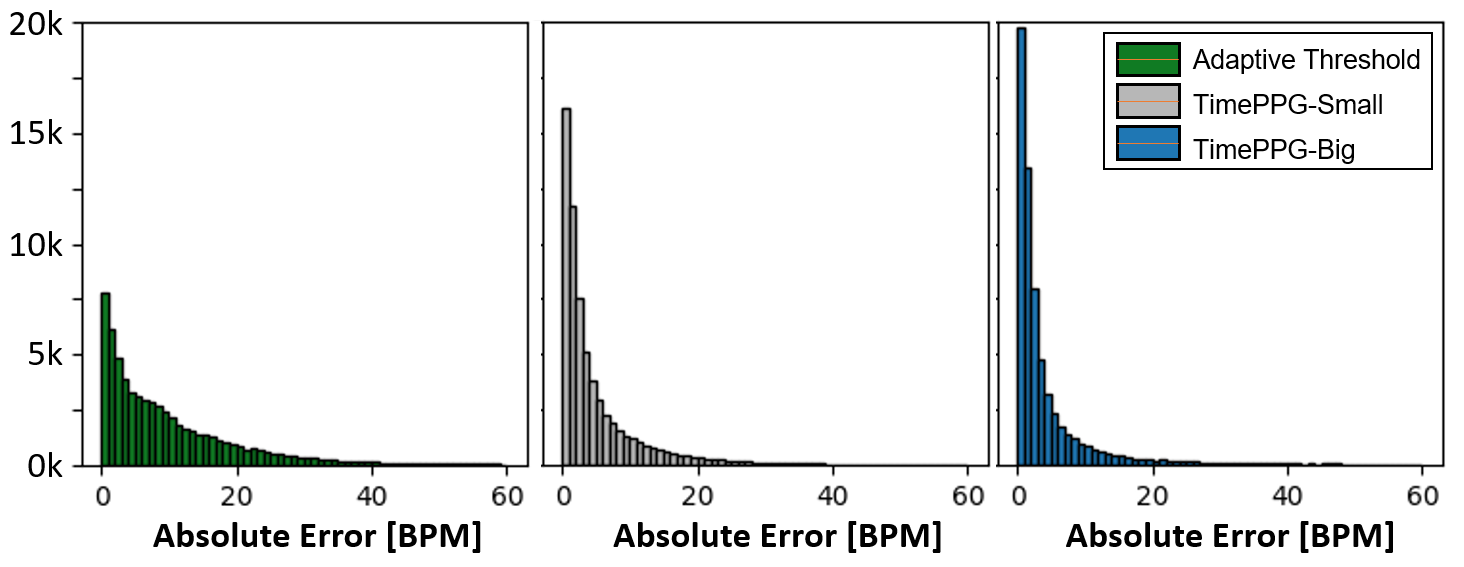}
  \caption{\minor{Distribution of the absolute errors associated with the three different base algorithms. For clarity of visualization, we cut the distribution at a max error of 60 BPM.}}
  \label{fig:errors}
  \end{figure}

As shown, our automatic design space exploration technique based on NAS spans more than one order of magnitude, both in terms of TCN parameters (5k-232k) and  OPs (0.1M-12.3M), despite starting from a single seed TCN.
The red triangles in the figure underline two relevant Pareto models from TimePPG, i.e., the most accurate and the smallest.
TimePPG-Big achieves the lowest overall MAE (4.88 BPM) while requiring around 232k parameters and 12.3M OPs.
TimePPG-Small is the smallest model found in our design space exploration and is obtained from MorphNet using a regularizer strength of 1e-5 and a pruning threshold of 0.01. This TCN has just 5.09k parameters (46$\times$ compression with respect to TimePPG-Big) and requires less than 100k OPs per inference to achieve a MAE of 5.63 BPM (0.75 increase). 

Finally, the figure also reports the results of the peak tracking algorithm of~\cite{RollingMean} (AT) as a red circle. Reporting this result in the figure is relevant, as AT will be later used as the ``small'' model in some ActPPG framework configurations.
Although this algorithm achieves a much higher overall MAE (11.0 BPM, shown not in scale in the graph), it has the advantage of not using the acceleration input signal, with a complexity of less than 3k OPs.
%
%
%

The first rows of Table \ref{tab:ActRes} report the accuracy results per subject obtained with AT, TimePPG-Small, and TimePPG-Big, \minor{whereas Table~\ref{tab:micro_deploy} reports the mean MAE across subjects and the number of operations per inference (OPs) of each model.
Furthermore, in Fig. \ref{fig:errors}, we report the error distribution of the three algorithms. Noteworthy, despite the fact that the MAE of the three algorithms is comprised between 4.88 BPM and 11.0 BPM, most of the errors are concentrated between 0 and 2 BPM. For instance, for TimePPG-Big, 30.5\% of the windows show an error $<$ 1 BPM and 51.3\% less than 2 BPM.}

\subsection{Movement Detector Results}

\begin{figure}[ht]
  \centering
  \includegraphics[width=0.6\columnwidth]{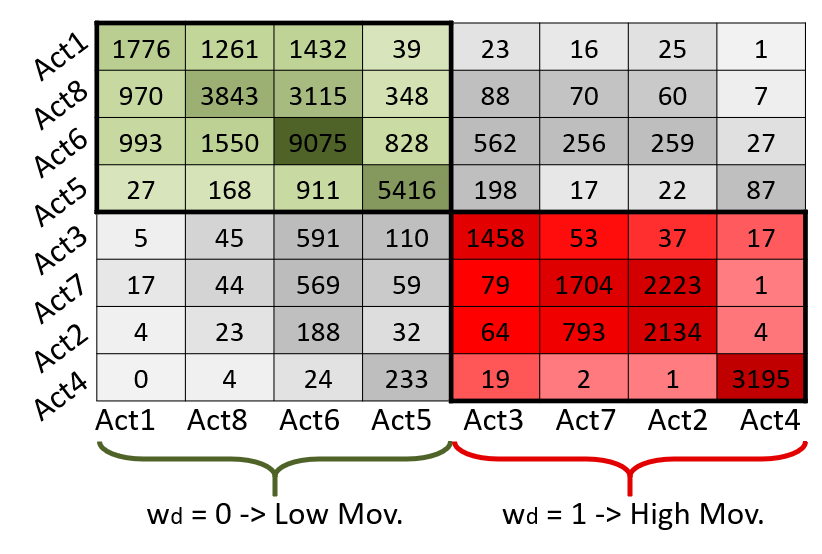}
  \caption{Confusion matrix over the eight activities of all 15 subjects of PPG-Dalia predicted by our random forest.}
  \label{fig:RF_confusion}
\end{figure}

As anticipated in Section~\ref{sec:ActPPG} our movement classifier is a random forest (RF) composed of 8 trees with a maximum depth of 5.
%
%
We use the mean value, energy, standard deviation, and the number of peaks of the 3D acceleration as input features, after a grid search on a more extended set of classical fatures.
%
As for HR estimation, we split the data using the cross-validation scheme proposed in~\cite{reiss2019deep} and report results as the average of all folds.
Figure~\ref{fig:RF_confusion} shows the confusion matrix of our random forest on the 8 activities contained in PPG-Dalia, where the activities have been sorted by increasing movement in the same way as for Figure~\ref{fig:acc_order}.
Considering eight separated classes, thus only excluding the ``transition'' phase of PPGDalia data, we achieve a micro-averaging accuracy $\nicefrac{TP+TN}{TS}$, of 60.6\%, (where $TP = $ true positives, $TN = $ true negatives, and $TS = $ total samples).
Noteworthy, most of the erroneous predictions mistake one class with one of its two nearest neighbors in terms of movement. Therefore, these errors do not cause problems for ActPPG, where the goal is only to distinguish ``low movement'' and ``high movement'' windows, as shown in the following.

Separating the activities in groups based on where the difference in terms of the average standard deviation of acceleration is the highest (i.e., between Act5 and Act3, see Figure~\ref{fig:acc_order}), we define the two macro classes low and high movement, separated by thicker lines in Figure~\ref{fig:RF_confusion}.
In distinguishing these 2 classes, the same random forest classifier achieves a 92.2\% accuracy.
When testing ActPPG, we verified that using either the random forest predictions or directly the ground truth activity labels to select the most appropriate HR estimator yields almost identical results in terms of MAE and number of ``big'' model calls, proving that this accuracy is sufficient for our task.
Therefore, we will employ the random forest predictions as a discriminator for the ActPPG framework in all the subsequent experiments.
\rev{Note that for the case of N$=$2 considered in this work, even a simple threshold on the average standard deviation of the 3-axial acceleration can achieve decent performance in separating low/high movement windows.}
\begin{table*}
\centering
\caption{\minor{In the first part of the table, the three algorithms used to test ActPPG. In the lower part, three configurations of ActPPG pairing two HR predictors and assigning Sitting, Working, Lunch and Driving activities to the smallest model in the pair. All results refer to PPG-Dalia dataset.}}
\label{tab:ActRes}
\renewcommand{\arraystretch}{1.2}
\small
\minor{
\begin{tabular}{|l|p{0.5cm}p{0.5cm}p{0.5cm}p{0.5cm}p{0.5cm}p{0.5cm}p{0.5cm}p{0.5cm}p{0.5cm}p{0.5cm}p{0.5cm}p{0.5cm}p{0.5cm}p{0.5cm}p{0.6cm}|}
Alg.                & \begin{tabular}[c]{@{}c@{}}S1\\Mean\\$\pm$sd\end{tabular}   & \begin{tabular}[c]{@{}c@{}}S2\\Mean\\$\pm$sd\end{tabular}    & \begin{tabular}[c]{@{}c@{}}S3\\Mean\\$\pm$sd\end{tabular}   & \begin{tabular}[c]{@{}c@{}}S4\\Mean\\$\pm$sd\end{tabular}   & \begin{tabular}[c]{@{}c@{}}S5\\Mean\\$\pm$sd\end{tabular}    & \begin{tabular}[c]{@{}c@{}}S6\\Mean\\$\pm$sd\end{tabular}    & \begin{tabular}[c]{@{}c@{}}S7\\Mean\\$\pm$sd\end{tabular}   & \begin{tabular}[c]{@{}c@{}}S8\\Mean\\$\pm$sd\end{tabular}   & \begin{tabular}[c]{@{}c@{}}S9\\Mean\\$\pm$sd\end{tabular}    & \begin{tabular}[c]{@{}c@{}}S10\\Mean\\$\pm$sd\end{tabular}  & \begin{tabular}[c]{@{}c@{}}S11\\Mean\\$\pm$sd\end{tabular}   & \begin{tabular}[c]{@{}c@{}}S12\\Mean\\$\pm$sd\end{tabular}  & \begin{tabular}[c]{@{}c@{}}S13\\Mean\\$\pm$sd\end{tabular}  & \begin{tabular}[c]{@{}c@{}}S14\\Mean\\$\pm$sd\end{tabular}  & \begin{tabular}[c]{@{}c@{}}S15\\Mean\\$\pm$sd\end{tabular}  \\ \hline \hline
\multicolumn{16}{|l|}{\textbf{TimePPG \& AT models: AT -1-, TimePPG-Small -2-, TimePPG-Big -3-}}\\\hline\hline
-1-                 & \begin{tabular}[c]{@{}c@{}}8.49\\ $\pm$9.7\end{tabular} & \begin{tabular}[c]{@{}c@{}}12.97\\ $\pm$10.4\end{tabular} & \begin{tabular}[c]{@{}c@{}}9.13\\ $\pm$8.7\end{tabular} & \begin{tabular}[c]{@{}c@{}}9.40\\ $\pm$8.5\end{tabular} & \begin{tabular}[c]{@{}c@{}}22.23\\ $\pm$19.1\end{tabular} &\begin{tabular}[c]{@{}c@{}} 17.55\\ $\pm$16.2\end{tabular} &\begin{tabular}[c]{@{}c@{}} 6.17\\ $\pm$6.5\end{tabular} & \begin{tabular}[c]{@{}c@{}}9.83\\ $\pm$9.6\end{tabular} & \begin{tabular}[c]{@{}c@{}}13.58\\ $\pm$11.0\end{tabular} & \begin{tabular}[c]{@{}c@{}}9.99\\ $\pm$14.1\end{tabular} & \begin{tabular}[c]{@{}c@{}}14.80\\ $\pm$15.2\end{tabular}& \begin{tabular}[c]{@{}c@{}}6.82\\ $\pm$6.6\end{tabular} & \begin{tabular}[c]{@{}c@{}}7.16\\ $\pm$7.7\end{tabular} & \begin{tabular}[c]{@{}c@{}}8.59\\ $\pm$8.0\end{tabular} & \begin{tabular}[c]{@{}c@{}}8.07\\ $\pm$9.5\end{tabular} \\
-2-            & \begin{tabular}[c]{@{}c@{}}5.57\\  $\pm$6.1\end{tabular} & \begin{tabular}[c]{@{}c@{}}4.57\\ $\pm$6.1\end{tabular}  & \begin{tabular}[c]{@{}c@{}}2.92\\ $\pm$3.5\end{tabular} & \begin{tabular}[c]{@{}c@{}}5.70\\ $\pm$6.9\end{tabular} & \begin{tabular}[c]{@{}c@{}}12.53\\ $\pm$18.3\end{tabular} & \begin{tabular}[c]{@{}c@{}}6.38\\ $\pm$9.5\end{tabular}  & \begin{tabular}[c]{@{}c@{}}3.12\\ $\pm$4.6\end{tabular} & \begin{tabular}[c]{@{}c@{}}6.76\\ $\pm$6.6\end{tabular} & \begin{tabular}[c]{@{}c@{}}7.60\\ $\pm$8.0\end{tabular}  & \begin{tabular}[c]{@{}c@{}}3.57\\ $\pm$5.6\end{tabular} & \begin{tabular}[c]{@{}c@{}}6.87\\ $\pm$9.3\end{tabular}  & \begin{tabular}[c]{@{}c@{}}8.47\\ $\pm$8.1\end{tabular} & \begin{tabular}[c]{@{}c@{}}2.69\\ $\pm$3.2\end{tabular} & \begin{tabular}[c]{@{}c@{}}3.47\\ $\pm$4.4\end{tabular} & \begin{tabular}[c]{@{}c@{}}4.34\\ $\pm$7.1\end{tabular} \\
-3-              & \begin{tabular}[c]{@{}c@{}}4.01\\ $\pm$4.9\end{tabular} & \begin{tabular}[c]{@{}c@{}}3.16\\ $\pm$4.2\end{tabular}  & \begin{tabular}[c]{@{}c@{}}2.27\\ $\pm$3.0\end{tabular} & \begin{tabular}[c]{@{}c@{}}4.62\\ $\pm$5.7\end{tabular} & \begin{tabular}[c]{@{}c@{}}14.96\\ $\pm$22.4\end{tabular} & \begin{tabular}[c]{@{}c@{}}4.28\\ $\pm$8.9\end{tabular}  & \begin{tabular}[c]{@{}c@{}}2.58\\ $\pm$3.7\end{tabular} & \begin{tabular}[c]{@{}c@{}}6.02\\ $\pm$6.9\end{tabular} & \begin{tabular}[c]{@{}c@{}}7.61\\ $\pm$8.8\end{tabular}  & \begin{tabular}[c]{@{}c@{}}2.89\\ $\pm$4.8\end{tabular} & \begin{tabular}[c]{@{}c@{}}4.79\\ $\pm$7.1\end{tabular}  & \begin{tabular}[c]{@{}c@{}}6.95\\ $\pm$8.8\end{tabular} & \begin{tabular}[c]{@{}c@{}}2.54\\ $\pm$2.9\end{tabular} & \begin{tabular}[c]{@{}c@{}}3.01\\ $\pm$4.0\end{tabular} & \begin{tabular}[c]{@{}c@{}}3.46\\ $\pm$6.3\end{tabular} \\ \hline \hline
\multicolumn{16}{|l|}{\textbf{ActPPG models} (e.g., 1 + 2 $=$ AT combined with TimePPG-Small)}\\\hline\hline
 1 + 2                 & \begin{tabular}[c]{@{}c@{}}5.95\\ $\pm$6.1\end{tabular} & \begin{tabular}[c]{@{}c@{}}7.64\\ $\pm$8.1\end{tabular}  & \begin{tabular}[c]{@{}c@{}}5.34\\ $\pm$6.1\end{tabular} & \begin{tabular}[c]{@{}c@{}}7.41\\ $\pm$7.3\end{tabular} & \begin{tabular}[c]{@{}c@{}}17.88\\ $\pm$17.5\end{tabular} & \begin{tabular}[c]{@{}c@{}}8.08 \\ $\pm$10.1\end{tabular} & \begin{tabular}[c]{@{}c@{}}4.18\\ $\pm$5.2\end{tabular} & \begin{tabular}[c]{@{}c@{}}8.12 \\ $\pm$7.9\end{tabular}& \begin{tabular}[c]{@{}c@{}}11.04\\ $\pm$10.1\end{tabular} & \begin{tabular}[c]{@{}c@{}}5.02 \\ $\pm$5.9\end{tabular}& \begin{tabular}[c]{@{}c@{}}8.70  \\ $\pm$9.6\end{tabular}& \begin{tabular}[c]{@{}c@{}}8.41 \\ $\pm$8.0\end{tabular}& \begin{tabular}[c]{@{}c@{}}3.94 \\ $\pm$4.2\end{tabular}& \begin{tabular}[c]{@{}c@{}}5.68 \\ $\pm$6.3\end{tabular}& \begin{tabular}[c]{@{}c@{}}5.39\\ $\pm$7.2\end{tabular}  \\
1 + 3                   & \begin{tabular}[c]{@{}c@{}}5.10 \\ $\pm$5.3\end{tabular}& \begin{tabular}[c]{@{}c@{}}6.59 \\ $\pm$7.4\end{tabular} & \begin{tabular}[c]{@{}c@{}}5.02\\ $\pm$6.1\end{tabular} & \begin{tabular}[c]{@{}c@{}}6.63\\ $\pm$6.4\end{tabular} & \begin{tabular}[c]{@{}c@{}}20.70\\ $\pm$20.8\end{tabular} & \begin{tabular}[c]{@{}c@{}}6.22 \\ $\pm$9.9\end{tabular} & \begin{tabular}[c]{@{}c@{}}3.65\\ $\pm$4.6\end{tabular} & \begin{tabular}[c]{@{}c@{}}8.00 \\ $\pm$8.1\end{tabular}& \begin{tabular}[c]{@{}c@{}}11.96\\ $\pm$10.7\end{tabular} & \begin{tabular}[c]{@{}c@{}}4.36\\ $\pm$5.5\end{tabular} & \begin{tabular}[c]{@{}c@{}}7.14 \\ $\pm$8.0\end{tabular} & \begin{tabular}[c]{@{}c@{}}7.91 \\ $\pm$8.6\end{tabular}& \begin{tabular}[c]{@{}c@{}}3.95 \\ $\pm$4.1\end{tabular}& \begin{tabular}[c]{@{}c@{}}5.32\\ $\pm$6.2\end{tabular} & \begin{tabular}[c]{@{}c@{}}4.79\\ $\pm$6.7\end{tabular} \\
2 + 3                & \begin{tabular}[c]{@{}c@{}}4.72 \\ $\pm$5.2\end{tabular}& \begin{tabular}[c]{@{}c@{}}3.52\\ $\pm$4.5\end{tabular}  & \begin{tabular}[c]{@{}c@{}}2.60\\ $\pm$3.4\end{tabular} & \begin{tabular}[c]{@{}c@{}}4.92\\ $\pm$5.7\end{tabular} & \begin{tabular}[c]{@{}c@{}}15.35\\ $\pm$22.2\end{tabular} & \begin{tabular}[c]{@{}c@{}}4.51 \\ $\pm$9.0\end{tabular} & \begin{tabular}[c]{@{}c@{}}2.59 \\ $\pm$3.7\end{tabular}& \begin{tabular}[c]{@{}c@{}}6.64 \\ $\pm$6.8\end{tabular}& \begin{tabular}[c]{@{}c@{}}8.52  \\ $\pm$9.1\end{tabular}& \begin{tabular}[c]{@{}c@{}}2.91\\ $\pm$4.8\end{tabular} & \begin{tabular}[c]{@{}c@{}}5.31 \\ $\pm$7.3\end{tabular} & \begin{tabular}[c]{@{}c@{}}7.97 \\ $\pm$8.7\end{tabular}& \begin{tabular}[c]{@{}c@{}}2.70 \\ $\pm$3.0\end{tabular}& \begin{tabular}[c]{@{}c@{}}3.11 \\ $\pm$4.1\end{tabular}& \begin{tabular}[c]{@{}c@{}}3.74 \\ $\pm$6.4\end{tabular} \\ \hline
\end{tabular}
}
\end{table*}
\rev{We use a RF not only to reach slightly more accurate results, but especially to make our framework easy to generalize to cases with N$>$2. Furthermore, the execution time and memory occupation of such a small RF is negligible compared to those of the HR tracking algorithms.}

\subsection{ActPPG: framework exploration}\label{sec:actppg_res}

\minor{The lowermost rows of Table~\ref{tab:ActRes} and Table~\ref{tab:micro_deploy} report the results of applying the ActPPG framework for combining two out of the three HR prediction models highlighted in Figure~\ref{fig:single_models} (AT, TimePPG-Small and TimePPG-Big).}
According to the previous movement detector results, these numbers are obtained by selecting the ``small'' HR predictor for ``low movement'' activities Act1 (Sitting), Act8 (Working), Act6 (Lunch), and Act5 (Driving), which span 51\% of the total data.

Using the two TimePPG networks as elements of ActPPG (last row), we obtain a low MAE of 5.27 BPM with a complexity which is approximately 50\% of that of the TimePPG-Big model alone on average. 
On the other hand, using the AT and TimePPG-Small allows reaching the lowest average complexity ($\sim$ 38.8 kOPs) at the cost of an increased MAE (7.52 BPM), yet significantly lower than the 10.99 BPM achieved by AT alone. 
\subsubsection{ActPPG: ablation study}
Here, we describe the two key elements that impact the performance and complexity of  ActPPG: the number of activities assigned to the small model and the percentage of low movement windows.

\begin{figure}[ht]
  \centering
  \includegraphics[width=0.6\columnwidth]{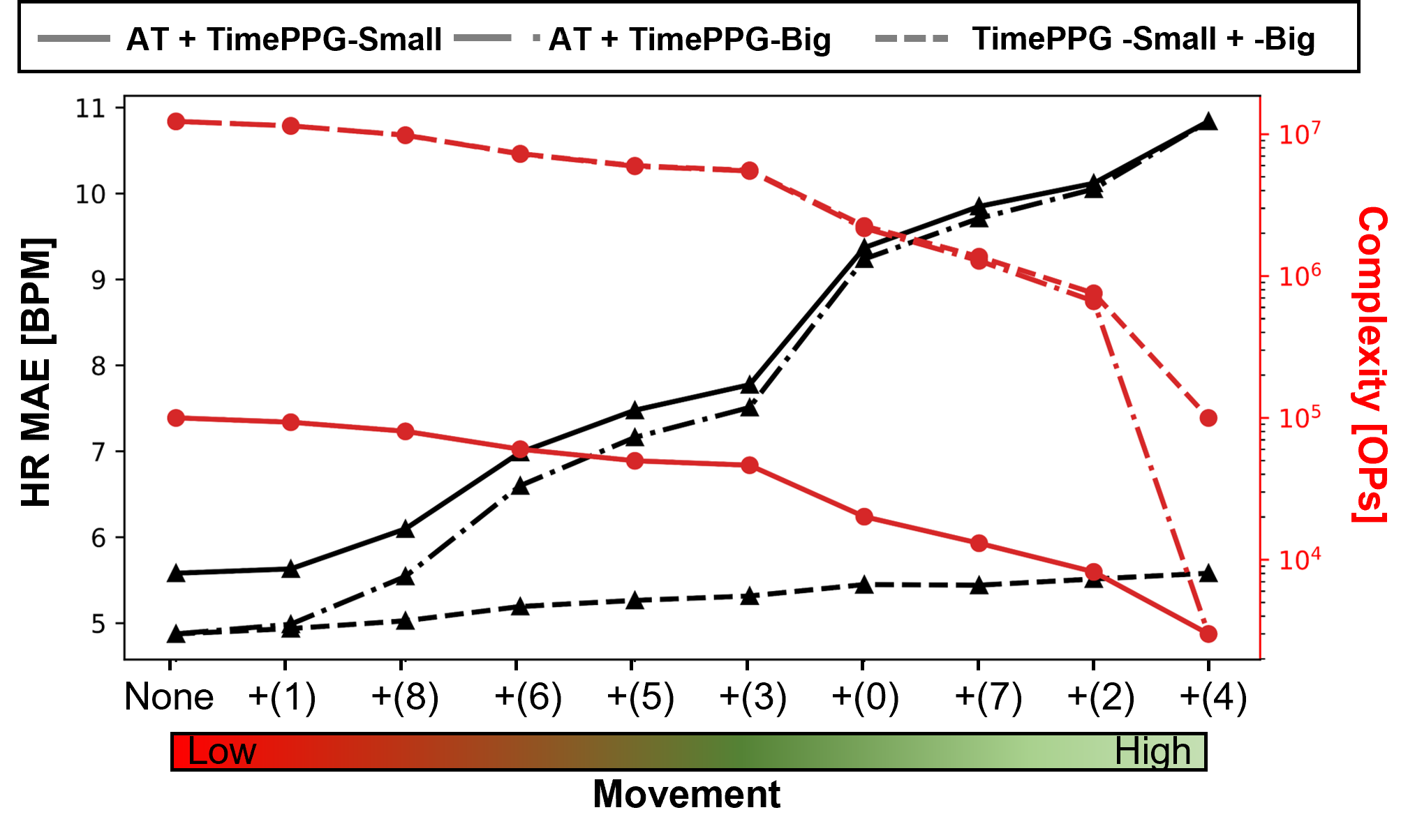}
  \caption{ActPPG framework performance and complexity as a function of the low/high movement activities separation. From left to right, more activities are tracked using the ``small'' model of the pair.}
  \label{fig:activities}
  \end{figure}

Figure~\ref{fig:activities} shows the impact of moving the boundary between low and high movement activities. Specifically, activity classes are progressively added to the ``low movement'' set one by one, in the order of Figure~\ref{fig:acc_order}.
In the two extreme cases (leftmost and rightmost points), we exclusively use the framework's big/small model, respectively.
Setting the threshold between activity 5 and 3, as shown in Figure~\ref{fig:RF_confusion}, we obtain an error increase of 1.89, 2.28, and 0.39 BPM for the combinations AT/TimePPG-Small, AT/TimePPG-Big, and TimePPB Small/Big with respect to the big models alone, while reducing the computational complexity by $\approx$ 50\%.
\begin{figure}[ht]
  \centering
  \includegraphics[width=0.6\columnwidth]{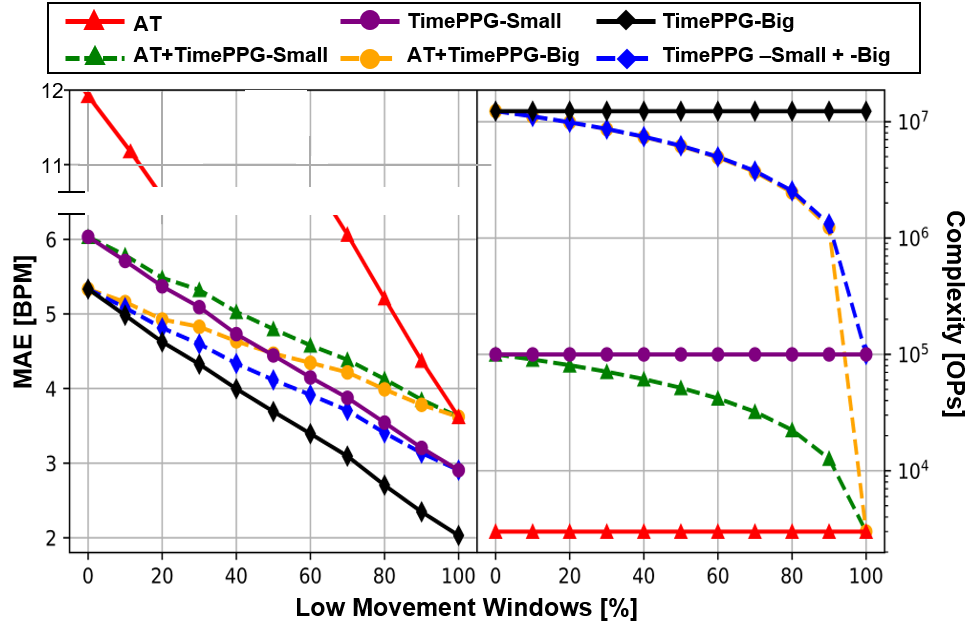}
  \caption{Complexity and MAE of all our models for a changing percentage of low movement windows processed.}
  \label{fig:moving_windows}
  \end{figure}

Notice that this threshold could also be changed dynamically at runtime, adapting the performance vs. complexity trade-off based on the system's status.
For instance, we could envision an ``energy-saving'' mode with a high threshold (i.e., most activities in the low-movement set), activated only when the wrist-worn device's battery charge is low. Similarly, a ``high-performance'' mode with a lower threshold could be activated manually by users who need more precise HR tracking for a short time. Finally, a ``normal-mode'' with the default threshold between Act5 and Act3 could be active in all other scenarios.

Figure~\ref{fig:moving_windows} analyzes the impact on our framework of users' daily routines, quantified as the number of low movement windows encountered by ActPPG.
The percentage of windows has been artificially changed by sampling from the original PPGDalia windows a different portion of the two classes, considering only activity 1 as low movement and the others as a high movement.
As the percentage of low movement windows increases, the algorithm's complexity always decreases, as expected, while the accuracy approximates the one of the smallest model.
\begin{table*}[ht]
\centering
\small
\renewcommand{\arraystretch}{1.2}
\begin{threeparttable}
\caption{Deployment of our models on the STM32L4R9AI. }
\label{tab:micro_deploy}
\rev{
\begin{tabular}{|llllllllll|}
\multicolumn{1}{l|}{}                                 & Ram [kB] & Flash [kB] & parameters                 & OPs     & Cycles [\#]                    & \multicolumn{1}{l|}{OPs/Cyc.} & \begin{tabular}[c]{@{}c@{}} Time\\{[}ms{]}\tnote{*}\end{tabular} & \begin{tabular}[c]{@{}c@{}} E.\\{[}mJ{]}\tnote{*}\end{tabular} & \begin{tabular}[c]{@{}c@{}} MAE\\{[}BPM{]}\tnote{+}\end{tabular} \\ \hline\hline
\multicolumn{10}{|l|}{\textbf{Single Models}}                                                                                                                                                                                                                                \\ \hline\hline
\multicolumn{1}{|l|}{AT (-1-)}                               & 4.09     & 0          & 0                          & 3k     & 100k                        & \multicolumn{1}{l|}{0.03}       & 1.25      & 0.017      &  10.99     \\
\multicolumn{1}{|l|}{TimePPG-Small (-2-)}                    & 7.68     & 18.54      & 5.09k                       & 77.63k    & 1.36M                       & \multicolumn{1}{l|}{0.057}      & 17.06     & 0.232      &  5.63      \\
\multicolumn{1}{|l|}{TimePPG-Small\tnote{\textdagger}} & 13.31    & 8.55       & 8.76k                       & 224.8k   & 1.52M                       & \multicolumn{1}{l|}{0.148}      & 19.02     & 0.259       & 5.60      \\
\multicolumn{1}{|l|}{TimePPG-Big (-3-)}                      & 129.64   & 902.21     & 232.6k & 12.27M & 103.16M & \multicolumn{1}{l|}{0.119}      & 1289.5    & 17.57 & 4.88      \\
\multicolumn{1}{|l|}{TimePPG-Big\tnote{\textdagger}}   & 34.06    & 884.26     & 902.2k & 33.3M & 104.14M                     & \multicolumn{1}{l|}{0.320}      & 1301.85   & 17.74      & 4.87      \\ \hline\hline
\multicolumn{10}{|l|}{\textbf{ActPPG models} (e.g., 1 + 2 $=$ AT combined with TimePPG-Small)}                                                                                                                                                                                                                                \\ \hline \hline
\multicolumn{1}{|l|}{1 + 2}                            & 7.68     & 20.04      & 6.63k                       & 38.81k    & 732.4k                        & \multicolumn{1}{l|}{0.053}       & 9.16      & 0.125      &  7.52      \\
\multicolumn{1}{|l|}{1 + 3}                            & 129.64   & 903.71     & 234.14k & 6.13M  & 51.63M                      & \multicolumn{1}{l|}{0.119}       & 645.38    & 8.796      &  7.16      \\
\multicolumn{1}{|l|}{2 + 3}                            & 129.64   & 922.25     & 239.2k                     & 6.17M  & 52.26M                      & \multicolumn{1}{l|}{0.118}       & 653.28    & 8.903      & 5.27      \\ \hline
\end{tabular}
\begin{tablenotes}
\item [\textdagger] With int8 quantization-aware training. Dilation reduced to one, with 0-padded filters to maintain the receptive field, to cope with toolchain limitations.
\item[*] Measured on the STM32L4R9AI with a frequency of 80MHz and a power consumption of 13.63 mW.
\item[+] Measured on the PPG-Dalia dataset.
\end{tablenotes}}
\end{threeparttable}
\end{table*}

In the rest of this sub-section, we focus mainly on the case of AT + TimePPG-Small, which mostly highlights this aspect.
Only executing AT, we have a constant low complexity of 3k OPs, but with quite bad performance results when high movement windows are frequent (leftmost part of the graph), reaching 12 BPM of MAE.
On the other hand, only executing TimePPG-small, we have a constant complexity of 77.63k OPs, but with a good MAE both in high movement (6 BPM) and low movement conditions (2.91 BPM).
In this case, the application of our framework (green dotted line of Figure~\ref{fig:moving_windows}) bounds the MAE in the [3.62, 6.03] BPM interval, avoiding bad performance in high movement situations while reducing the complexity of up to 33$\times$ during low movement situations compared to the execution of TimePPG-Small alone.
In other words, ActPPG can bring the advantages of both small and big models by employing the appropriate model in every situation, thus minimally increasing the MAE (green vs. purple curves) while significantly reducing the mean complexity and therefore the energy consumption.

\subsection{Algorithms Deployment}
\label{subsec:HW_deployment}

This section details the results obtained deploying some of our solutions (both individual models and ActPPG combinations using the default activity separation threshold fixed between Act5 and Act3) on the embedded platform described in Section~\ref{sec:hardware_setup}.
\rev{We report the memory occupation, the inference time, and the energy consumption for tracking an 8 seconds window in Table~\ref{tab:micro_deploy}.
The results refer to the deployment on the STM32L4R9AI MCU, clocked at 80 MHz, with a power consumption of 13.63 mW.
Given that \textit{int8} quantized deep neural networks reach iso-accuracies compared to their \textit{float32} counterparts in literature \cite{zanghieri2019robust}, we show both the results of the original float32 and of int8 models (\textdagger).
We used NEMO~\cite{conti2020technical} to re-train our models with a quantization-aware training, achieving negligible accuracy loss compared to float32.
For model deployment, we used CUBE.AI 5.1.2, which supports both full-precision and quantized models.
However, the toolchain does not yet support dilation on quantized models. Therefore, we had to manually extend the convolution filters interleaving them with 0s to replicate the effect of dilation\footnote{\rev{For instance, a filter with $C_{in}=1$, $K=3$, $d = 2$ and values \{3,-7,43\} had to be changed to $K=5$ with values \{3, 0, -7, 0, 43\} }}.
Further energy and latency reductions (at iso-accuracy) could be achieved once the toolchain adds support for quantized dilation.}

\rev{The simple AT algorithm of~\cite{RollingMean}  achieves a very short latency of 1.25 ms, which is 13.65$\times$/1041.5$\times$ lower than the ones of our two extreme TimePPG outputs, i.e., TimePPG-Small and TimePPG-Big, which reach 17.06 ms and 1289.5 ms in their float32 deployment. 
When quantized, both TCNs show a similar latency ($\sim$ 2 ms slower) despite the higher MACs/cycle achieved (0.148 vs 0.057 for TimePPG-Small and 0.320 vs. 0.119 for TimePPG-Big). This is due to the the larger total number of MACs executed (2.7$\times$ / 2.9$\times$ more respectively) because of the 0-interleaving described above.
Using quantized models also reduces the memory occupation (although less than what would be achieved without the additional 0s), which is beneficial especially if the MCU has to store multiple networks.
Notably, all the algorithms meet the requirement of having a total execution time $<$ 2s and can be considered suitable for real-time HR tracking.}

\rev{
%
%
In terms of energy, AT consumes just 17 $\mu$J, while TimePPG-Small and TimePPG-Big (deployed in float) consume 0.232 mJ and the 17.57 mJ respectively.
Note in particular that both the inference time and the energy consumption of TimePPG-Big are  two orders of magnitude larger than those of TimePPG-Small,  due to the large number of computations performed by this TCN.
Combining this model with TimePPG-Small through the ActPPG framework, we reduce its total energy consumption by $\sim$ 50\%, from 17.57 mJ to 8.903 mJ, with a MAE increase of just 0.39 BPM. Similarly, also the ActPPG combination of AT and TimePPG-Small yields relevant energy consumption reductions compared to the latter, with a significantly lower error than the former. All ActPPG results in Table~\ref{tab:micro_deploy} also include the memory and latency overheads due to the RF-based movement detector.}

\begin{figure}[ht]
  \centering
  \includegraphics[width=0.6\columnwidth]{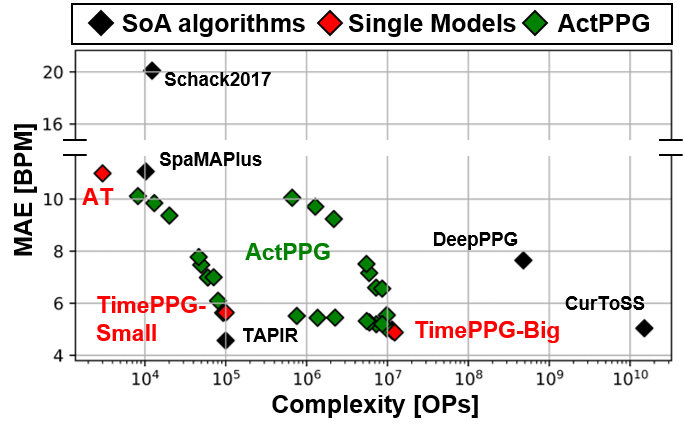}
  \caption{Comparison with SoA algorithms in the Complexity vs. MAE space. Red points represent the three models employed in ActPPG, while green points represents Pareto-optimal solutions of ActPPG obtained by changing the threshold between low and high movement activities. For deep learning methods, we consider MACs as OPs. }
  \label{fig:SOA}
  \end{figure}
  
\subsection{State-of-the-art comparison}\label{sec:soa_compare}
\begin{table}[ht]
\centering
\caption{Comparison of TimePPG models with state-of-the-art methods. The table reports the mean MAE over subjects $\pm$ the MAE standard deviation over subjects, when available. For our work, we report TimePPG-Small and TimePPG-Big obtained for PPG-Dalia (top) and SPC2015 (bottom). Abbreviations: Mem.: memory, B: Byte, E.: energy.}\label{tab:soa_compare}
\renewcommand{\arraystretch}{1.2}
\begin{tabular}{l|ll|llll|}
                                                                      & \multicolumn{2}{c|}{MAE [BPM]}    & \multicolumn{4}{c|}{Deployment}                  \\ \hline
\multicolumn{1}{|l|}{Algorithm}                                       & SPC2015       & PPG-Dalia             & [Ops/MACs] & Parameters [\#] & Mem. [B]   & E.[mJ] \\ \hline\hline
\multicolumn{7}{|l|}{\textbf{Classical Algorithms} }                                                                                                                       \\ \hline \hline
\multicolumn{1}{|l|}{TROIKA, 2014 \cite{troika2014}}                      & 2.34 $\pm$ 0.82 & n.a. & n.a.  & n.a.       & n.a.     & n.a.        \\
\multicolumn{1}{|l|}{JOSS, 2015 \cite{joss2015}}                      & 1.28 $\pm$ 2.61 & n.a.  & 15.0B  & n.a.       & n.a.     & n.a.        \\
\multicolumn{1}{|l|}{SpaMaPlus$^1$, 2016 \cite{spama2016}}                & 4.25 $\pm$ 5.9  & 11.06 $\pm$ 4.8      & 10.3k  & n.a.       & n.a.     & n.a.        \\
\multicolumn{1}{|l|}{Schack, 2017   \cite{schack2017computationally}} & 2.91 $\pm$ 4.6  & 20.45 $\pm$ 7.1      & 12.3k  & n.a.       & n.a.     & n.a.        \\
\multicolumn{1}{|l|}{TAPIR, 2020 \cite{huang2020robust}}              & 2.5 $\pm$ 1.2   & 4.57 $\pm$ 1.4       & 100k   & n.a.       & n.a.     & n.a.        \\
\multicolumn{1}{|l|}{Arunkumar, 2020   \cite{arunkumar2020robust}}    & 1.03 $\pm$ 1.49 & n.a.                 & n.a.       & n.a.       & n.a.     & n.a.        \\
\multicolumn{1}{|l|}{CurToSS, 2020 \cite{zhou2020heart}}              & 2.2             & 5.0 $\pm$ 2.8        & 15.0B  & n.a.       & n.a.     & n.a.        \\
\multicolumn{1}{|l|}{AT} & 11.92 $\pm$ 5.39 & 10.99 $\pm$ 4.46 & 3k        & 0       & 0   & 17 $\mu$J        \\ \hline\hline
\multicolumn{7}{|l|}{\textbf{Deep Learning Algorithms}}                                                                                                                  \\ \hline\hline
\multicolumn{1}{|l|}{CNN, 2019 \cite{reiss2019deep}}         & 4.0 $\pm$ 5.4   & 7.65 $\pm$ 4.2       & 480M       & 60M    & 240M & n.a.        \\
\multicolumn{1}{|l|}{CNN constr., 2019   \cite{reiss2019deep}}    & n.a.            & 9.99 $\pm$ 5.9       & 380k       & 26k        & 26k      & n.a.        \\
\multicolumn{1}{|l|}{PPGNet, 2019 \cite{shyam2019ppgnet}}             & 3.36 $\pm$ 4.1  & n.a.                 & n.a.       & 765k       & n.a.     & n.a.        \\ 
\multicolumn{1}{|l|}{CorNET, 2019 \cite{cornet2019}}                  & 4.67 $\pm$ 3.71 & n.a                  & 20M        & 250k       & 1M       & n.a.        \\
\multicolumn{1}{|l|}{Bin CorNET, 2020   \cite{rocha2020binary}}    & 6.78 $\pm$ 5.29 & n.a                  & 20M        & 250k       & 260k     & 56.1 $\mu$J$^2$ \\
\multicolumn{1}{|l|}{Chung, 2020 \cite{chung2020deep}}                & 1.46$^3$            & n.a.                 & 17M        & 3.3M       & n.a.     & n.a.        \\
\multicolumn{1}{|l|}{\minor{DeepHeart, 2021 \cite{deepheart2021}}}                &\minor{ 1.61 }           & \minor{n.a.  }               & \minor{1101M   }     & \minor{4.3M }      & \minor{n.a.  }   & \minor{n.a.  }      \\
\hline\hline
\multicolumn{7}{|l|}{\textbf{Our work}}                                                                                                                  \\ \hline\hline
\multicolumn{1}{|l|}{TimePPG-Small} & 4.82 $\pm$ 2.75 & 5.63 $\pm$ 2.63 & \begin{tabular}[c]{@{}c@{}}77.6k \\ 1.77M\end{tabular} & \begin{tabular}[c]{@{}c@{}}5k \\ 11k\end{tabular}   & \begin{tabular}[c]{@{}c@{}}18.54k \\ 43.31k\end{tabular}  & \begin{tabular}[c]{@{}c@{}}232 $\mu$J$^4$  \\ 3.54 mJ$^4$\end{tabular} \\
\multicolumn{1}{|l|}{TimePPG-Big} & 3.27 $\pm$ 2.04 & 4.88 $\pm$ 3.23 & \begin{tabular}[c]{@{}c@{}}12.3M \\ 16.4M\end{tabular} & \begin{tabular}[c]{@{}c@{}}232k\\ 168k\end{tabular}   & \begin{tabular}[c]{@{}c@{}}902.2k \\ 658.8k\end{tabular}  & \begin{tabular}[c]{@{}c@{}}17.57 mJ$^4$  \\ 23.98 mJ$^4$ \end{tabular} \\\hline
\multicolumn{7}{l}{$^1$  Original paper reported MAE = 0.89 BPM on SPC2015, but not using leave-one-subject-out validation.}\\  
\multicolumn{7}{l}{$^2$  Deployed on ASIC.}\\  
\multicolumn{7}{l}{\begin{tabular}[c]{@{}l@{}}$^3$  The SPC2015 dataset has been used only for training and thus the accuracy is not computed using \\leave-one-subject-out validation.\end{tabular}}\\ 
\multicolumn{7}{l}{$^4$  Deployed on general purpose MCU STM32L4R9AI.}\\
\end{tabular}
\end{table}

Figure~\ref{fig:SOA} shows a full comparison of our models (green and red diamonds) with state-of-the-art algorithms (black ones), including both classical and deep learning solutions, benchmarked on the PPG-Dalia dataset.
Green diamonds represent ActPPG results obtained varying the threshold between low and high movement activities as in Figure~\ref{fig:activities}.
Our models cover the space from 4.88 BPM to 10.99 BPM of MAE, with a complexity that spans from 12 MOps to 0.003 MOPs.
As shown in the graph, both our individual models and our adaptive framework obtain results comparable with the state-of-the-art, sometimes significantly reducing the complexity compared to other approaches with similar accuracy. For instance, TimePPG-Big achieves a lower MAE than CurToSS (4.88 BPM vs. 5.0 BPM) but has 1000$\times$ lower complexity.
Further, compared to DeepPPG (the previous state-of-the-art deep learning model tested on PPG-Dalia), TimePPG-Small achieves better MAE (5.63 BPM vs. 7.65 BPM), while reducing the size and the OPs by 12000$\times$ and 4800$\times$, respectively.

Table \ref{tab:soa_compare} further strengthens this conclusion, by comparing our individual TimePPG models to state-of-the-art techniques tested on both the SPC2015 and the PPG-Dalia dataset\footnote{\rev{To allow a direct comparison to previous models, we also collected a Pareto-optimal frontier of TCN models on the SPC2015 12 subjects dataset always using the leave-one-subject-out validation and the parameters proposed in \cite{reiss2019deep}. As for PPG-Dalia dataset, ``small'' and ``big'' models are extracted from the Pareto front and reported in Table \ref{tab:SoA}.}}. 
Compared to previous deep learning algorithms \cite{shyam2019ppgnet,cornet2019,chung2020deep, deepheart2021}, our approach replaces more expensive layers such as LSTMs with dilated 1D convolutions, allowing a strong reduction in the number of parameters and in the computational time. 
As shown, both ``extreme'' TimePPG models are characterized by less parameters and MACs compared to the state-of-the-art, while achieving comparable or lower MAE on both the benchmark datasets.
For instance, TimePPG-Small has 3.8$\times$/4800$\times$ lower MACs and 5.2$\times$/12000$\times$ lower parameters compared to others deep learning approaches, while achieving better performance on the PPG-Dalia dataset (5.63 vs. 7.65 BPM). Similarly, TimePPG-Big obtains comparable performance to PPGNet on the SPC2015 dataset (3.27 vs 3.36 BPM) with 4.6$\times$ less parameters.
\minor{DeepHeart \cite{deepheart2021} outperforms our networks in terms of MAE on the SPC2015 dataset, obtaining a 1.61 BPM. However, this network can not be deployed on constrained MCUs such as the ones considered in this work, given that it requires 1101M MACs and has 4.3M of parameters, not fitting the typical $<$ 1 MB memory of microcontrollers.}

\rev{While our TCNs outperform previous deep learning approaches, some model-driven methods still obtain slightly better results on both datasets, reaching as low as 4.57 BPM of MAE on PPG-Dalia and 1.03 BPM on SPC2015.
However, these methods have different shortcomings.
%
%
Firstly, they perform poorly both when tested on different data compared to those on which they are tuned, or under slight parameters modifications. Indeed, most model-based algorithms (e.g., ~\cite{spama2016,schack2017computationally}) tuned for SPC2015 perform much worse on PPG-Dalia, confirming that the tuning of their parameters is critical and not trivial. The few exceptions (e.g., ~\cite{huang2020robust,zhou2020heart}) are only due to the fact that the parameters of those algorithms have been tuned on both datasets \textit{simultaneously}, and without leave-one-subject-out validation.
Secondly, none of the model-based algorithms (except the trivial AT) has been deployed for real-time execution on an edge device, neither a general purpose MCU or a dedicated ASIC/FPGA. Vice versa, they are only tested offline and without consideration of real-time execution or energy consumption.}

Moreover, as previously discussed in Section~\ref{sect:arch_opt_improv}, our models are strongly impaired by subject 5, \minor{whose large HR values are rarely encountered in training data and are therefore badly predicted by data-driven approaches.}
In fact, if we apply the additional fine-tuning step described in Section~\ref{sect:arch_opt_improv}, TimePPG-Big (and consequently also the ActPPG models based on it) outperforms also classical methods, reaching 3.84 BPM of MAE on PPG-Dalia. 
\minor{Due to the smaller size of the dataset, and to the way in which it was collected (with the first 25\% of all subjects' records containing only low-movement activities), the proposed fine-tuning strategy does not yield similar improvements on SPC2015. However, note that as explained in Section~\ref{sect:arch_opt_improv}, the purpose of this experiment was not to reduce the MAE per se, but just to show that our proposed TCNs could achieve even better performance given the availability of a larger and more comprehensive dataset.}

\subsection{\minor{Limitations}}

\minor{In this section, we summarize the limitations of the proposed method, most of which have been addressed individually elsewhere in the manuscript. Similarly to all other PPG-based HR tracking algorithms (both classic and deep learning ones), also our TimePPG TCNs are negatively affected by intense MAs. This is discussed in Sections~\ref{sec:ActPPG} and \ref{sec:actppg_res}, and shown clearly by Figures~\ref{fig:Act_order} and \ref{fig:moving_windows}, which demonstrate the better performance achieved by individual models for low-movement inputs. However, the ActPPG framework goes exactly in the direction of addressing this intrinsic limitations of individual models, providing a mechanism to achieve good HR tracking performance at the minimum possible cost, automatically adapting to the amount of MAs.

Since we rely on pre-collected datasets (PPG-Dalia and SPC2015) for the training and evaluation of our models, the effectiveness of the proposed method with different PPG sensing hardware, e.g., not wrist-mounted, or in general providing an output signal with different characteristics, cannot be assessed. Once again, this limitation is shared with the majority of previous literature, which similarly provides results only on pre-collected, open-source, data. 
%
%
Our plans for future works go in the direction of addressing this limitation, through the collection of larger and more heterogeneous datasets in terms of activities, HR ranges, and sensing hardware.

This also relates with the last limitation of our method, i.e., the fact that our TCNs need large amounts of training data to be accurate. This is a common problem of data-driven algorithms, and in particular deep learning ones, and was discussed extensively in Sections~\ref{sect:arch_opt_ft} and \ref{sec:soa_compare}. In particular, the need for large datasets is demonstrated by the much better results (in relation with the state-of-the-art) achieved by our method on PPG-Dalia, compared to the smaller SPC2015. The positive effect of fine-tuning on Dalia also confirms this point. Nonetheless, even without fine-tuning, our TCNs almost match the state-of-the-art on Dalia, and significantly outperform all previous data-driven solutions. This demonstrates that, thanks to the use of small and optimized neural networks, our method is less affected than previous deep learning ones by this problem. Moreover, this limitation can also be seen as unexpressed potential. In fact, even PPG-Dalia is a very small dataset by today's deep learning standards. So, the fact that we are competitive on it is an indication that, given the availability of more data, our method could achieve even better performance.}

\section{Conclusions}
\label{sec:conclusions}

HR tracking is increasingly executed on wearable devices both in clinical contexts and in daily lives. 
However, the advanced HR tracking algorithms proposed in the literature are tested only offline and not always applicable in real-time on constrained devices. To the best of our knowledge, there is a lack of research results focusing on the deployment of advanced algorithms on MCU-class systems and on their optimization from the point of view of energy consumption. 

In this work, we have presented a two-fold contribution that goes in this direction.
%
%
First, we have presented TimePPG, a set of deep learning models for HR tracking based on TCN architectures, which achieve excellent performance with a limited number of parameters. All models are obtained, starting from a single seed using an automatic NAS approach.
Second, we have proposed a run-time framework, ActPPG, to select among multiple HR prediction algorithms based on the movement intensity.
The intuition behind this solution comes from the fact that a higher movement intensity causes more relevant MAs and is, therefore, more difficult to process.
Both TimePPG and ActPPG obtain accurate results that are competitive with the state-of-the-art, while executing in real-time on a low-power MCU. 
On the PPGDalia dataset, the largest TimePPG model, coupled with post-processing and fine-tuning, reduces the MAE to as low as 3.84 BPM.
Moreover, ActPPG generates a rich set of Pareto solutions that achieve MAEs in the range [4.88, 10.99] BPM and span four orders of magnitude in complexity, between 12.3 MOPs and 0.003 MOPs.

\section{Acknowledgments}
Support was received from the Hasler Foundation under project 18082, and the EU’s Horizon 2020 Research and Innovation Program through the project MNEMOSENE under Grant 780215.

\bibliographystyle{ACM-Reference-Format}
\bibliography{references}

\end{document}